\documentclass[minimal]{informs2}

\usepackage{endnotes}

\usepackage{amssymb}
\usepackage{amsmath}
\usepackage{color}
\usepackage{graphicx}
\usepackage{comment}
\usepackage{algorithm}
\usepackage{algorithmic}
\usepackage[margin=1in]{geometry}

\usepackage{natbib}
 \bibpunct[, ]{(}{)}{,}{a}{}{,}%
 \def\bibfont{\small}%
\TheoremsNumberedThrough     
\ECRepeatTheorems

\EquationsNumberedThrough    

\usepackage{subfig}
\begin{document}



\RUNTITLE{Optimal Pricing in Networks with Externalities}

\TITLE{Optimal Pricing in Networks with Externalities}

\ARTICLEAUTHORS{%
\AUTHOR{Ozan Candogan}
\AFF{Department of Electrical Engineering and Computer Science\\Massachusetts Institute of Technology, MA, Cambridge, MA 02139, \EMAIL{candogan@mit.edu}} 
\AUTHOR{Kostas Bimpikis}
\AFF{Operations Research Center and Department of Electrical Engineering and Computer Science\\Massachusetts Institute of Technology, MA, Cambridge, MA 02139, \EMAIL{kostasb@mit.edu}} 
\AUTHOR{Asuman Ozdaglar}
\AFF{Department of Electrical Engineering and Computer Science\\Massachusetts Institute of Technology, MA, Cambridge, MA 02139, \EMAIL{asuman@mit.edu}} 
} 

\ABSTRACT{We study the optimal pricing strategies of a monopolist selling a divisible good (service) to consumers that are embedded in a social network. A key feature of our model is that consumers experience a (positive) \textit{local network effect}. In particular, each consumer's usage level depends directly on the usage of her \textit{neighbors} in the social network structure. Thus, the monopolist's optimal pricing strategy may involve offering discounts to certain agents, who have a \textit{central} position in the underlying network. Our results can be summarized as follows. First, we consider a setting where the monopolist can offer individualized prices and derive an explicit characterization of the optimal price for each consumer as a function of her network position. In particular, we show that it is optimal for the monopolist to charge each agent a price that is proportional to her \textit{Bonacich centrality} in the social network. In the second part of the paper, we discuss the optimal strategy of a monopolist that can only choose a single uniform price for the good and derive an algorithm polynomial in the number of agents to compute such a price. Thirdly, we assume that the monopolist can offer the good in two prices, full and discounted, and study the problem of determining which set of consumers should be given the discount. We show that the problem is NP-hard, however we provide an explicit characterization of the set of agents that should be offered the discounted price. Next,  we describe an approximation algorithm for finding the optimal set of agents. 
We show that if the profit is nonnegative under any feasible price allocation,
the algorithm guarantees at least 88$~\%$ of the optimal profit. Finally, we highlight the value of network information by comparing the profits of a monopolist that does not take into account the network effects when choosing her pricing policy to those of a monopolist that uses this information optimally.}

\AREAOFREVIEW{Revenue Management.} 
\maketitle

\newcommand{\todo}[1]{\vspace{5 mm}\par \noindent \marginpar{\textsc{ToDo}}
\framebox{\begin{minipage}[c]{0.95 \columnwidth}
\tt #1 \end{minipage}}\vspace{5 mm}\par}
\section{Introduction} 
\label{sec:introduction}

Inarguably social networks, that describe the pattern and level of interaction of a set of agents\footnote{We use the terms ``agent'' and ``consumer'' interchangeably.}, are instrumental in the propagation of information and act as conduits of influence among its members. Their importance is best exemplified by the overwhelming success of online social networking communities, such as Facebook and Twitter. The ubiquity of these internet based services, that are built around social networks, has made possible the collection of vast amounts of data on the structure and intensity of social interactions. The question that arises naturally is whether firms can intelligently use the available data to improve their business strategies.

In this paper, we focus on the question of using the potentially available data on network interactions to improve the pricing strategies of a seller, that offers a divisible good (service). A main feature of the products we consider is that they exhibit a local (positive) \textit{network effect}: increasing the usage level of a consumer has a positive impact on the usage levels of her peers. As concrete examples of such goods, consider online games (e.g., World of Warcraft, Second Life) and social networking tools and communities (e.g., online dating services, employment websites etc.). More generally, the local network effect can capture \textit{word of mouth} communication among agents: agents typically form their opinions about the quality of a product based on the information they obtain from their peers.

How can a monopolist exploit the above network effects and maximize her revenues? In particular, in such a setting it is plausible that an optimal pricing strategy may involve favoring certain agents by offering the good at a discounted price and subsequently exploiting the positive effect of their usage on the rest of the consumers. At its extreme, such a scheme would offer the product for free to a subset of consumers hoping that this would have a large positive impact on the purchasing decisions of the rest. Although such strategies have been used extensively in practice, mainly in the form of ad hoc or heuristic mechanisms, 
the  available data enable companies to effectively target the agents to maximize that impact. 

The goal of the present paper is to characterize optimal pricing strategies as a function of the underlying social interactions in a stylized model, which features consumers that are embedded in a given social network and influencing each other's decisions. In particular, a monopolist first chooses a pricing strategy and then consumers choose their usage levels, so as to maximize their own utility. We capture the local positive network effect by assuming that a consumer's utility is increasing in the usage level of her peers. We study three variations of the baseline model by imposing different assumptions on the set of available pricing strategies, that the monopolist can implement. 

First, we allow the monopolist to set an individual price for each of the consumers. We show that the optimal price for each agent can be decomposed into three components: a fixed cost, that does not depend on the network structure, a markup and a discount. 
Both the markup and the discount are proportional to the \textit{Bonacich centrality} of the agent's neighbors in the social network structure, which is a sociological measure of network influence. 
The Bonacich centrality measure, introduced by \cite{Bonacich:1987}, 
can be computed as the stationary distribution of a random walk on the underlying network structure. Hence, the agents with the highest centrality are the ones that are visited by the random walk most frequently.
Intuitively,  agents get a  discount proportional to the amount they influence their peers to purchase the product, and  they receive a markup if they are strongly influenced by other agents in the network.
Our results provide an economic foundation for this sociological measure of influence.

Perfect price differentiation is typically hard to implement. Therefore, in the second part of the paper we study a setting, where the monopolist offers a single uniform price for the good. Intuitively, this price might make the product unattractive for a subset of consumers, who end up not purchasing, but the monopolist recovers the revenue losses from the rest of the consumers. We develop an algorithm that finds the optimal single price in time polynomial in the number of agents. The algorithm considers different subsets of the consumers and finds the optimal price provided  that only the consumers in the given subset  purchase a positive amount of the good. First, we show that given a subset $S$ we can find the optimal price $p_{S}$ under the above constraint in closed form. Then, we show that we only need to consider a small number of such subsets. In particular, we rank the agents with respect to a weighted centrality index and at each iteration of the algorithm we drop the consumer with the smallest such index and let $S$ be the set of remaining consumers.

Finally, we consider an intermediate setting, where the monopolist can choose one of a small number of prices for each agent. For exposition purposes, we restrict the discussion to two prices, \textit{full} and \textit{discounted}. We show that the resulting problem, i.e., determining the optimal subset of consumers to offer the discounted price, is NP-hard \footnote{The hardness result can be extended to the case of more than two prices.}. We also provide an approximation algorithm that recovers (in polynomial time) at least 88$~\%$ of the optimal revenue.

To further highlight the importance of network effects, we compare the profits of a monopolist that ignores them when choosing her pricing policy to those of a monopolist that exploits them optimally. We are able to provide a concise characterization of this discrepancy as a function of the level of interaction between the agents. Informally, the value of information about the network structure increases with the level of asymmetry of interactions among the agents.

As mentioned above, a main feature of our model is the positive impact of a consumer's purchasing decision to the purchasing behavior of other consumers. This effect, known as \textit{network externality}, is extensively studied in the economics literature (e.g., \cite{Farrell:1985}, \cite{Katz:1986}). However, the network effects in those studies are of \textit{global} nature, i.e., the utility of a consumer depends directly on the behavior of the whole set of consumers. In our model, consumers interact directly only with a subset of agents. Although interaction is local for each consumer, her utility may depend on the global structure of the network, since each consumer potentially interacts indirectly with a much larger set of agents than just her peers. 

Given a set of prices, our model takes the form of a \textit{network game} among agents that interact locally. A recent series of papers studies such games, e.g., \cite{Ballester:2006}, \cite{Bramoulle:2007}, \cite{Corbo:2007}, \cite{Galeotti:2009}. A key modeling assumption in \cite{Ballester:2006}, \cite{Bramoulle:2007} and \cite{Corbo:2007}, that we also adopt in our setting, is that the payoff function of an agent takes the form of a linear-quadratic function. Ballester et al. in  \cite{Ballester:2006} were the first to note the linkage between Bonacich centrality and Nash equilibrium outcomes in a single stage game with local payoff complementarities. Our characterization of optimal prices when the monopolist can perfectly price differentiate is reminiscent of their results, since prices are inherently related to the Bonacich centrality of each consumer. However, both the motivation and the analysis are quite different, since ours is a two-stage game, where a monopolist chooses prices to maximize her revenue subject to equilibrium constraints. Also, \cite{Bramoulle:2007} and \cite{Corbo:2007} study a similar game to the one in \cite{Ballester:2006} and interpret their results in terms of public good provision. A number of recent papers (\cite{Campbell:2009}, \cite{Galeotti:2010} and \cite{Sundararajan:2007}) have a similar motivation to ours, but take a completely different approach: they make the assumption of limited knowledge of the social network structure, i.e., they assume that only the degree distribution is known, and thus derive optimal pricing strategies  that depend on this first degree measure of influence of a consumer. In our model, we make the assumption that the monopolist has complete knowledge of the social network structure and, thus, obtain qualitatively different results: the degree is not the appropriate measure of influence but rather prices are proportional to the Bonacich centrality of the agents. On the technical side, note that assuming more global knowledge of the network structure increases the complexity of the problem in the following way: if only the degree of an agent is known, then essentially there are as many different \textit{types} of agents as there are different degrees. This is no longer true when more is known: then, two agents of the same degree may be of different type because of the difference in the characteristics of their neighbors, and therefore, optimal prices charged to agents may be different.

Finally, there is a recent stream of literature in computer science, that studies a set of algorithmic questions related to marketing strategies over social networks. Kempe et al. in \cite{Kempe:2003} discuss optimal \textit{network seeding} strategies over social networks, when consumers act myopically according to a pre-specified rule of thumb. In particular, they distinguish between two basic models of diffusion: the \textit{linear threshold model}, which assumes that an agent adopts a behavior as soon as adoption in her neighborhood of peers exceeds a given threshold and \textit{independent cascade model}, which assumes that an adopter \textit{infects} each of her neighbors with a given probability. The main question they ask is finding the optimal set of initial adopters, when their number is given, so as to maximize the eventual adoption of the behavior, when consumers behave according to one of the diffusion models described above. They show that the problem of \textit{influence maximization} is NP-hard and provide a greedy heuristic, that achieves a solution, that is provably within 63 $\%$ of the optimal. 

Closest in spirit with our work, is \cite{Hartline:2008}, which discusses the optimal marketing strategies of a monopolist. Specifically, they assume a general model of influence, where an agent's willingness to pay for the good is given by a function of the subset of agents that have already bought the product, i.e., $u_i: 2^{V} \rightarrow {\mathbb{R}}_{+}
$, where $u_i$ is the willingness to pay for agent $i$ and $V$ is the set of consumers. They restrict the monopolist to the following set of \textit{marketing strategies}: the seller visits the consumers in some sequence and makes a take-it-or-leave-it offer to each one of them. Both the sequence of visits as well as the prices are chosen by the monopolist. They provide a dynamic programming algorithm that outputs the optimal pricing strategy for a symmetric setting, i.e., when the agents are ex-ante identical (the sequence of visits is irrelevant in this setting). Not surprisingly the optimal strategy offers discounts to the consumers that are visited earlier in the sequence and then extracts revenue from the rest. The general problem, when agents are heterogeneous, is NP-hard, thus they consider approximation algorithms. They show, in particular, that \textit{influence-and-exploit} strategies, that offer the product for free to a strategically chosen set $A$, and then offer the myopically optimal price to the remaining  agents provably achieve a constant factor approximation of the optimal revenues under some assumptions on the influence model. However, this paper does not provide a qualitative insight on the relation between optimal strategies and the structure of the social network. In contrast, we are mainly interested in characterizing the optimal strategies as a function of the underlying network. 

The rest of paper is organized as follows. Section \ref{sec:model} introduces the model. In Section \ref{sec:consumption} we begin our analysis by characterizing the usage level of the consumers at equilibrium given the vector of prices chosen by the monopolist. In Section \ref{sec:pricing} we turn attention to the pricing stage (first stage of the game) and characterize the optimal strategy for the monopolist under three different settings: when the monopolist can perfectly price discriminate (Subsection \ref{subsec:perfect}), when the monopolist chooses a single uniform price for all consumers (Subsection \ref{subsec:single}) and finally when the monopolist can choose between two exogenously given prices, the full and the discounted (Subsection \ref{subsec:two}). In Section \ref{sec:value}, we compare the profits of a monopolist that has no information about the network structure (and thus chooses her pricing strategy as if consumers did not interact with one another) with those of a monopolist that has full knowledge over the network structure and can perfectly price discriminate consumers. Finally, we conclude in Section \ref{sec:conclusions}. To ease exposition of our results, we decided to relegate the proofs to the Appendix.
\section{Model}
\label{sec:model}The society consists of a set $\mathcal{I}=\{1,\ldots, n\}$ of agents embedded in a social network represented by the adjacency matrix $G$. The $ij$-th entry of $G$, denoted by $g_{ij}$, represents the \textit{strength} of the influence of agent $j$ on $i$. We assume that $g_{ij} \in [0,1]$ for all $i, j$ and we normalize $g_{ii}=0$ for all $i$.
A monopolist introduces a \textit{divisible} good in the market and chooses a vector $\mathbf{p}$ of prices from the set of allowable \textit{pricing strategies} $\mathbf{P}$. In its full generality, $\mathbf{p} \in \mathbf{P}$ is simply a mapping from the set of agents to $ \mathbb{R}^n$, i.e., 
$\mathbf{p}: \mathcal{I} \rightarrow \mathbb{R}^n.$
In particular, $\mathbf{p}(i)$ or equivalently $p_i$
is the price that the monopolist  offers to agent $i$ for one unit of the divisible good. Then, the agents choose the amount of the divisible good they will purchase at the announced price. Their utility is given by an expression of the following form: 
$$u_i (x_i,\mathbf{x_{-i}},p_i) = f_i(x_i) + x_i h_i\left(G, \mathbf{x_{-i}}\right) - p_i x_i,$$
where $x_i \in [0,\infty)$ is the amount of the divisible good that agent $i$ chooses to purchase. Function $f_i: [0,\infty) \rightarrow \mathbb{R}$ represents the utility that the agent obtains from the good, assuming that there are no network externalities, and $p_i x_i$ is the amount agent $i$ is charged for its consumption. The function
 $h_i: [0,1]^{n \times n} \times [0,\infty)^{n-1} \rightarrow [0, \infty)$ is used to capture the utility the agent  obtains  due to the positive network effect (note the explicit dependence on the network structure). 

 We next  describe the two-stage \textit{pricing-consumption} game, which models the interaction between the agents and the monopolist:\\
 \textbf{Stage 1 (Pricing) :} The monopolist chooses the pricing strategy $\mathbf{p}$, so as to maximize profits, i.e., 
 $\max_{\mathbf{p} \in \mathbf{P}}  \sum_{i} p_i x_i - c x_i,$
 where $c$ denotes the marginal cost of producing a unit of the good and $x_i$ denotes the amount of the good agent $i$ purchases in the second stage of the game.
 \\
 \textbf{Stage 2 (Consumption) :} Agent $i$ chooses to purchase $x_i$ units of the good, so as to maximize her utility given the prices chosen by the monopolist and $\mathbf{x}_{-i}$, i.e., 
 \[x_i \in \arg\max_{y_i \in [0,\infty)}u_i (y_i,\mathbf{x_{-i}},p_i).\]
  We are interested in the \textit{subgame perfect} equilibria of the two-stage pricing-consumption game.
  
  For  a fixed vector of prices ${\mathbf p}=[p_i]_i$ chosen by the monopolist,
the equilibria of the second stage game,  referred to as the consumption equilibria, are defined as  follows:
  \begin{definition}[Consumption Equilibrium] For a given vector of prices $\mathbf{p}$, a vector $\mathbf{x}$ is a consumption equilibrium if, for all $i\in {\cal I}$,
      \[
      ~{x_i} \in \arg\max_{y_i \in [0,\infty)}u_i (y_i,\mathbf{x_{-i}}, p_i).~
      \]
      We denote the set of consumption equilibria at a given price vector $\mathbf{p}$ by $C[\mathbf{p}]$.
      \end{definition}
   We begin our analysis by the second stage (the consumption subgame) and then discuss the optimal pricing policies for the monopolist given that agents purchase according to the consumption equilibrium of the subgame defined by the monopolist's choice of prices.
 
\section{Consumption Equilibria}
\label{sec:consumption}

For the remainder of the paper, we assume that the payoff function of agent $i$ takes the following quadratic form:
\begin{equation} \label{eq:quadUtilGen}
u_{i}(x_i, \mathbf{x_{-i}},p_i)= a_ix_i - b_ix_i^2 + x_i \cdot \sum_{j \in \{1,\cdots,n\}} g_{ij}\cdot  x_{j} -p_i x_i,
\end{equation}
where the first two terms represent the utility agent $i$ derives from consuming $x_i$ units of the good irrespective of the consumption of her peers, the third term represents the (positive) network effect of her social group and finally the last term is the cost of usage. The quadratic form of the utility function allows for tractable analysis, but also serves as a good second-order approximation of the broader class of concave payoffs.

For a given vector of prices $\mathbf{p}$, we denote by ${\cal G}=\{ {\cal I}, \{u_i\}_{{i\in {\cal I}}}, {[0, \infty)}_{i\in {\cal I}}  \}$  the second stage game where the set of players is $\cal I$,    each player $i\in {\cal I}$  chooses her strategy (consumption level) from the set $[0,\infty)$, and her the utility function, $u_i$ has the form in \eqref{eq:quadUtilGen}.
The following assumption ensures that in this game the optimal consumption level of each agent is bounded.
\begin{assumption} \label{ass:concavityCond}
For all $i\in {\cal I}$, $b_i> \sum_{j\in {\cal I}} g_{ij}$. 
\end{assumption}

The necessity of Assumption \ref{ass:concavityCond} is evident from the following example: assume that the adjacency matrix, which represents the level of influence among agents, takes the following simple form: $g_{ij}=1$ for all $i,j$ such that $i\neq j$, i.e., $G$ represents a complete  graph with unit weights. Also, assume that $0<b_i=b < n-1$ and $0<a_i=a$ for all $i \in \mathcal{I}$. It is now straightforward to see that given any vector of prices $\mathbf{p}$ and assuming that $x_i=x$ for all $i\in {\cal I}$, 
the payoffs of all agents go to infinity as $x\rightarrow \infty$.
Thus, if Assumption \ref{ass:concavityCond} does not hold,  in the consumption game,  consumers may choose to unboundedly increase their usage  irrespective of the vector of prices.

Next, we study the second stage of the game defined in Section \ref{sec:model} under Assumption \ref{ass:concavityCond}, and we characterize the equilibria of the consumption game among the agents for vector of prices $\mathbf{p}$. In particular, we show that the equilibrium is unique and we provide a closed form expression for it. To express the results in a compact form, we define the  vectors
$\mathbf{x},\mathbf{a}, \mathbf{p}\in \mathbb{R}^n$ such that 
${\mathbf x}=[x_i]_i$, ${\mathbf a}=[a_i]_i$, ${\mathbf p}=[p_i]_i$.  We also define matrix $\Lambda \in \mathbb{R}^{n\times n}$ as:
\begin{equation*}\notag
\Lambda_{i,j}= \left\{
\begin{aligned}
2b_i& \qquad \mbox{if $i= j$} \\
0& \qquad \mbox{otherwise.} 
\end{aligned}
\right.
\end{equation*}

Let $\beta_i(\mathbf{x_{-i}})$ denote the best response of agent $i$, when the rest of the agents choose consumption levels represented by the vector $\mathbf{x_{-i}}$. From \eqref{eq:quadUtilGen} it follows that:
\begin{equation}
\label{eq:br_basic}
\beta_i(\mathbf{x_{-i}})=\max\left\{\frac{a_i-p_i}{2 b_i}  + \frac{1}{2 b_i} \sum_{j\in {\cal I}} g_{ij} x_j , 0 \right\}.
\end{equation}
Our first result shows that the equilibrium of the consumption game is unique for any price vector. 
\begin{theorem} \label{eq:uniqueEq}
Under Assumption \ref{ass:concavityCond}, the game ${\cal G}=\{ {\cal I}, \{u_i\}_{{i\in {\cal I}}}, {[0, \infty)}_{i\in {\cal I}}  \}$ has a unique equilibrium.
\end{theorem}

Intuitively, Theorem \ref{eq:uniqueEq}   follows from the fact that increasing one's consumption incurs a positive externality on her peers, which further implies that the game involves strategic complementarities and therefore the equilibria are ordered. The proof exploits this monotonic ordering to show that the equilibrium is actually unique.

We conclude this section, by characterizing the unique equilibrium of $\cal G$.
Suppose that $\mathbf{x}$ is this equilibrium, and $x_i>0$ only for $i\in S$. 
Then, it follows that
\begin{equation} \label{eq:xiFOC_unique}
x_i = \beta_i(\mathbf{x_{-i}})=\frac{a_i-p_i}{2 b_i}  + \frac{1}{2 b_i} \sum_{j\in {\cal I}} g_{ij} x_j =\frac{a_i-p_i}{2 b_i}  + \frac{1}{2 b_i} \sum_{j\in { S}} g_{ij} x_j 
\end{equation}
for all $i\in S$.
Denoting by $\mathbf{x}_S
$ the vector of all $x_i$ such that $i\in S$, and defining the vectors $\mathbf{a}_S
$, $\mathbf{b}_S
$, $\mathbf{p}_S$ and the matrices $G_S$, $\Lambda_S$ similarly, equation \eqref{eq:xiFOC_unique} can be rewritten as
\begin{equation}\label{eq:xSFOC_unique}
\Lambda_S  \mathbf{x}_S
 = \mathbf{a}_S
-\mathbf{p}_S + G_S \mathbf{x}_S
.
\end{equation}
Note that Assumption \ref{ass:concavityCond} holds for the graph restricted to the agents in $S$, hence $I - \Lambda_S^{-1} G_S$ is invertible (cf. Lemma \ref{lem:invertible} in the Appendix). Therefore, \eqref{eq:xSFOC_unique} implies that
\begin{equation} \label{eq:eqS_unique}
\mathbf{x}_S
=    (\Lambda_S - G_S)^{-1} (\mathbf{a}_S
-\mathbf{p}_S).
\end{equation}
Therefore, the unique equilibrium of the consumption game takes the following form:
\begin{equation} \label{eq:eq_characterization}
\begin{array}{ll}
\mathbf{x}_S
=   (\Lambda_S - G_S)^{-1} (\mathbf{a}_S
-\mathbf{p}_S),\\
\mathbf{x_{\mathcal{I}-S}} = \mathbf{0},
\end{array}
\end{equation}
for some subset $S$ of the set of agents $\mathcal{I}$. This characterization suggests that consumptions of players (weakly) decrease with the prices. The following lemma, which is used in the subsequent analysis, formalizes this fact.
\begin{lemma} \label{lem:monotoneDec}
Let $\mathbf{x(\mathbf{p})}$ denote the unique consumption equilibrium in the game where each player $i\in {\cal I}$ is offered the price $p_i$. Then, $x_i(\mathbf{p})$ is weakly decreasing in $\mathbf{p}$ for all $i\in{\cal I}$, i.e, if $\hat{\mathbf{p}}_j \geq \mathbf{p}_j$ for all $j\in {\cal I}$ then $x_i(\hat{\mathbf{p}}) \leq x_i(\mathbf{p})$. 
\end{lemma}

\section{Optimal Pricing}
\label{sec:pricing}

In this section, we turn attention to the first stage of the game, where a monopolist sets the vector of prices. We distinguish between three different scenarios. In the first subsection, we assume that the monopolist can \textit{perfectly price discriminate} the agents, i.e., there is no restriction imposed on the prices.
In the second subsection, we consider the problem of choosing a single uniform price, while in the third we allow the monopolist to choose between two exogenous prices, $p_L
$ and $p_H
$, for each consumer. In our terminology, in the first case $\mathbf{P}=\mathbb{R}^{|\cal{I}|}$, in the second  $\mathbf{P}=\{(p,\cdots,p)\}$, for $p\in [0, \infty)$ and finally in the third $\mathbf{P}=\{p_L
,p_H
\}^{|\cal{I}|}$.
\subsection{Perfect Price Discrimination}
\label{subsec:perfect}
\noindent For the remainder of the paper, we make the following assumption, which ensures that, even in the absence of any network effects, the monopolist would find it optimal to charge individual prices low enough, so that all consumers purchase a positive amount of the good.
\begin{assumption} 
\label{ass:all_buy}
For all $i\in {\cal I}$, $a_i>c$.
\end{assumption}
\noindent Given Assumption \ref{ass:all_buy}, we are now ready to state Theorem \ref{theo:optPrice}, that provides a characterization of the optimal prices. We denote the vector of all 1's by $\mathbf{1}$.
\begin{theorem} \label{theo:optPrice}
Under Assumptions \ref{ass:concavityCond} and \ref{ass:all_buy}, the optimal prices are given by
\begin{equation}
\label{eq:eq_optPrice}
{\mathbf p}= {\mathbf a} -(\Lambda - G) \left(\Lambda - \frac{G+G^T}{2}\right)^{-1} \frac{{\mathbf a}-c{\mathbf 1}
}{2}.
\end{equation}
\end{theorem}
The following corollary is an immediate consequence of Theorem \ref{theo:optPrice}.
\begin{corollary} \label{cor:symmetricNetwork}
Let Assumptions \ref{ass:concavityCond} and \ref{ass:all_buy} hold. Moreover, assume that the interaction matrix $G$ is symmetric. Then, the optimal prices satisfy
\begin{equation}
\notag
{\mathbf p}= \frac{{\mathbf a}+c{\mathbf 1}
}{2}, \notag
\end{equation}
i.e., the optimal prices do not depend on the network structure.
\end{corollary}

This result implies that when players affect each other in the same way, i.e., when the interaction matrix $G$ is symmetric, then the graph topology has no effect on the optimal prices. 

To better  illustrate the effect of the network structure on prices we next consider a special setting, in which agents are symmetric in a sense defined precisely below and they differ only in terms of their network position.
\begin{assumption}\label{ass:symPlayers}
Players are symmetric, i.e., $a_i= a_0$, $b_i=b_0$ for all $i \in \cal{I}$.
\end{assumption}

We next provide the definition of Bonacich Centrality (see also \cite{Bonacich:1987}). We use this definition to obtain an alternative characterization of the optimal prices.
\begin{definition}[Bonacich Centrality]
For a network with (weighted) adjacency matrix $G$ and scalar $\alpha$, the Bonacich centrality vector of parameter $\alpha$ is given by
${\cal K} (G,\alpha)=(I-\alpha G)^{-1} {\mathbf 1}$ provided that $(I-\alpha G)^{-1}$ is well defined and nonnegative.
\end{definition}

\begin{theorem} \label{theo:optPriceSymPfirst}
Under Assumptions \ref{ass:concavityCond}, \ref{ass:all_buy} and \ref{ass:symPlayers}, the vector of optimal prices is given by
\begin{equation*}
\begin{aligned}
{\mathbf p} &= \frac{{a_0}+{c
}}{2} {\mathbf 1} + \frac{{a_0}-{c
}}{8b_0} G {\cal K} \left(\frac{G+G^T}{2}, \frac{1}{2b_0} \right)-\frac{{a_0}-{c
}}{8b_0} G^T {\cal K} \left(\frac{G+G^T}{2},\frac{1}{2b_0}\right).
\end{aligned}
\end{equation*}
\end{theorem}

The network $\frac{G+G^T}{2}$ is the average interaction network, and it represents the average interaction between pairs of agents in network $G$. Intuitively, the centrality $ {\cal K} \left(\frac{G+G^T}{2},\frac{1}{2b_0}\right)$ measures how ``central'' each agent is with respect to the average interaction network. 

The optimal prices in Theorem \ref{theo:optPriceSymPfirst} have three components. The first component can be thought of as a nominal price, which is charged to all agents irrespective of the network structure. The second term is a markup that the monopolist can impose on the price of consumer $i$ due to the utility the latter derives from her peers. Finally, the third component can be seen as a discount term, which is offered to a consumer, since increasing her consumption increases the consumption level of her peers. Theorem \ref{theo:optPriceSymPfirst} suggests that it is  optimal to give each agent a markup proportional to the utility she derives from the central agents. In contrast,  prices offered to the agents should be discounted proportionally to their influence on central agents.
Therefore, it follows that the agents which pay the most favorable prices are the ones, that \textit{influence} highly central agents.

Note that if Assumption \ref{ass:symPlayers} fails, then Theorem \ref{theo:optPriceSymPfirst} can be modified to relate the optimal prices to  centrality measures in the underlying graph. In particular, the price structure is still as given  in \eqref{eq:priceStructure},   but when the parameters $\{a_i\}$ and $\{b_i\}$ are not  identical, the discount and markup terms are proportional to a  weighted version of the Bonacich centrality measure, defined below.
\begin{definition}[Weighted Bonacich Centrality]
For a network with (weighted) adjacency matrix $G$, diagonal matrix $D$ and weight vector $\mathbf{v}$, the weighted Bonacich centrality vector is given by
$\tilde{\cal K} (G,D,\mathbf{v})=(I- G D)^{-1} {\mathbf{v}}$ provided that $(I- GD)^{-1}$ is well defined and nonnegative.
\end{definition}
We next characterize the optimal prices in terms of the weighted Bonacich centrality measure.
\begin{theorem} \label{theo:optPriceSymP2}
Under Assumptions \ref{ass:concavityCond} and \ref{ass:all_buy}   the vector of optimal prices is given by
\begin{equation*}
\begin{aligned}
{\mathbf p} &= \frac{{\mathbf{a}}+{c  {\mathbf 1}
}}{2} + G  \Lambda^{-1} \tilde{\cal K} \left( \tilde{G}, \Lambda^{-1}, \tilde{\mathbf{v}} \right)-
G^T \Lambda^{-1} \tilde{\cal K} \left( \tilde{G}, \Lambda^{-1}, \tilde{\mathbf{v}} \right),
\end{aligned}
\end{equation*}
where $\tilde{G}= \frac{G+G^T}{2}$ and $\tilde{\mathbf{v}}= \frac{{\mathbf{a}}-{c\mathbf{1}
}}{2} $.
\end{theorem}
\subsection{Choosing a Single Uniform Price}
\label{subsec:single}

In this subsection we  characterize the equilibria of the pricing-consumption game, when the monopolist can only set a single uniform price, i.e., $p_i=p_0$ for all $i$. Then, for any fixed $p_0$,  the payoff function of agent $i$ is given by
\begin{equation}\notag
u_{i}(x_i, \mathbf{x_{-i}},p_i)= a_ix_i - b_ix_i^2 + x_i \cdot \sum_{j \in \{1,\cdots,n\}} g_{ij}\cdot  x_{j} -p_i x_i,
\end{equation}
and the payoff function for the monopolist is given by
\begin{equation*}
\begin{array}{ll}\max_{p_0 \in [0, \infty)} & (p_0 - c)\sum_{i} x_i \\
s.t. & \mathbf{x} \in C[\mathbf{p}_0],
\end{array}
\end{equation*}
where $\mathbf{p}_0=(p_0,\cdots,p_0)$. Note that Theorem \ref{eq:uniqueEq} implies that even when the monopolist offers a single price, the consumption game has a unique equilibrium point.  The next lemma states that the consumption of each agent decreases monotonically in the price.

\begin{lemma} \label{lem:monotoneDec2}
Let $\mathbf{x(p_0)}$ denote the unique equilibrium in the game where $p_i=p_0$ for all $i$. Then, $x_i(\mathbf{p_0})$ is weakly decreasing in $p_0$ for all $i\in{\cal I}$ and strictly decreasing for all $i$ such that $x_i(\mathbf{p_0})>0$.
\end{lemma}

Next, we introduce  the notion of the centrality gain.

\begin{definition}[Centrality Gain]
In a network with (weighted) adjacency matrix $G$, for any diagonal matrix $D$ and weight vector $\mathbf{v}$, the centrality gain of agent $i$ is defined as
\begin{equation}
\notag
H_i(G,D,\mathbf{v})=\frac{\tilde{\cal K}_i(G,D,\mathbf{v})}{\tilde{\cal K} _i(G,D,\mathbf{1})}.
\end{equation}
\end{definition}

The following theorem provides a characterization of the consumption vector at equilibrium as a function of the single uniform price $p$.
\begin{theorem}
\label{thm:singleprice}
%
Consider game 
$\bar{\cal G}=\{ {\cal I}, \{u_i\}_{i\in {\cal I}}, {[0,\infty)}_{i\in {\cal I}}  \}$, 
and define 
\[D_1=\arg\min_{i\in{\cal I}} H_i \left(  G , \Lambda^{-1}, \mathbf{a} \right) \quad \textrm{ and } \quad p_1=  \min_{i\in{\cal I}} H_i \left(   G , \Lambda^{-1},\mathbf{a} \right).\] 
Moreover, let $I_k={\cal I}- \cup_{i=1}^{k} D_i$
 and define 
 \[D_k=\arg\min_{i\in{ I_k}} H_i \left(   G_{I_k}  ,\Lambda^{-1}_{I_k},\mathbf{a}_{I_k} \right) \quad \textrm{ and } \quad p_k=   \min_{i\in{ I_k}} H_i \left(  G_{I_k},\Lambda^{-1}_{I_k},\mathbf{a}_{I_k} \right),\]
 for $k\in \{2, 3 \dots n\}$.
Then,
\begin{enumerate}
\item[(1)] $p_k$ strictly increases in $k$.
\item[(2)] 
Given a $p$ such that $p<p_1$, all agents purchase a positive amount of the good, i.e., $x_i(p)>0$  for all $i\in{\cal I}$, where $\mathbf{x(p)}$ denotes the unique consumption equilibrium at price $p$. If $k\geq1$, and $p$ is such that
$p_{k}\leq p \leq p_{k+1}$, then $x_i(p)>0$ if and only if $i\in I_k$. Moreover, the corresponding consumption levels are given as in \eqref{eq:eq_characterization}, where $S=I_k$. 
\end{enumerate}
\end{theorem}

\noindent Theorem \ref{thm:singleprice} also suggests a polynomial time algorithm for computing the optimal uniform price $p_{opt}$. Intuitively, the algorithm  sequentially removes consumers with the lowest centrality gain and computes the optimal price for the remaining consumers under the assumption that the price is low enough so that only these agents purchase a positive amount of the good at the associated consumption equilibrium. In particular, using Theorem \ref{thm:singleprice}, it is possible to identify the set of agents who purchase a positive amount of the good for  price ranges $[p_k,p_{k+1}]$, $k\in \{1, \dots\}$. Observe that  given a set of players, who purchase a positive amount of the good, the equilibrium consumption levels can be obtained in closed form as a linear function of the offered price, and, thus, the profit function of the monopolist takes a quadratic form in the price.  It follows that for each price range, the maximum profit can be found by solving a quadratic optimization problem. 
Thus, Theorem \ref{thm:singleprice} 
suggests Algorithm \ref{alg:singleprice} for finding the optimal single uniform price $p_{opt}$.
\begin{algorithm}
\caption{: Compute the optimal single uniform price $p_{opt}$}
\label{alg1}
\begin{algorithmic}
\label{alg:singleprice}
\item[] \textbf{STEP 1.}  Preliminaries:
\begin{itemize}
\item[-] Initialize the set of \textit{active} agents: $S:=\cal{I}$.
\item[-] Initialize $k=1$ and $p_0=0$, $p_1=\min_{i\in{\cal{I}}} H_i(G_{\cal{I}},\Lambda_{\cal I}^{-1},\mathbf{a_{\cal{I}}})$
\item[-] Initialize the monopolist's revenues with $Re_{opt}=0$ and $p_{opt}=0$.
\end{itemize}
\item[] \textbf{STEP 2.}  
\begin{itemize}
\item[-] Let $\hat{p} =  \frac{\mathbf{1}^T (\Lambda_S - G_S)^{-1} \mathbf{a}_S  -  c \mathbf{1}^T (\Lambda_S - G_S)^{-1} \mathbf{1}}{\mathbf{1}^T  \left( (\Lambda_S - G_S)^{-1} + (\Lambda_S - G_S^T)^{-1} \right) \mathbf{1}} $
\item[-] \textbf{IF } $\hat{p} \geq p_{k}$, let $p=p_{k}$.
\item[] \textbf{ELSE IF } $\hat{p} \leq p_{k-1}$, let $p=p_{k-1}$ \textbf{ELSE } $p=\hat{p}$.
\item[-] $Re=(p-c)\mathbf{1}^{T}\cdot (\Lambda_S - G_S)^{-1} (\mathbf{a}_S-p \mathbf{1})$. 
\item[-] \textbf{IF } $Re > Re_{opt}$ \textbf{THEN } $Re_{opt} = Re$ and $p_{opt} = p$.
\item[-] $D=\arg\min_{i\in{ S}} H_i(G_{S},\Lambda^{-1}_S,\mathbf{a}_{S})$ and $S:=S-D$.
\item[-] Increase $k$ by 1 and let $p_k= \min_{i\in{ S}} H_i(G_{S},\Lambda^{-1}_S,\mathbf{a}_{S}).$
\item[-] Return to \textbf{STEP 2} if $S \neq \emptyset$ \textbf{ELSE}  \textbf{Output} $p_{opt}.$
\end{itemize}
\end{algorithmic}
\end{algorithm}

The algorithm solves a series of subproblems, where the monopolist is constrained to choose a price $p$ in a given interval $[p_k, p_{k+1}]$ with appropriately chosen endpoints. In particular, from Theorem \ref{thm:singleprice}, we can choose those endpoints, so as to ensure that only a particular set $S$ of agents purchase a positive amount of the good. In this case, the consumption at price $p$ is given by $ (\Lambda_S - G_S)^{-1} (\mathbf{a}_S-p \mathbf{1})$ and the profit of the monopolist  is equal to $(p-c)\mathbf{1}^{T} (\Lambda_S - G_S)^{-1} (\mathbf{a}_S-p \mathbf{1})$. The maximum of this profit function is achieved at $\hat{p} =  \frac{\mathbf{1}^T (\Lambda_S - G_S)^{-1} \mathbf{a}_S  -  c \mathbf{1}^T (\Lambda_S - G_S)^{-1} \mathbf{1}}{\mathbf{1}^T  \left( (\Lambda_S - G_S)^{-1} + (\Lambda_S - G_S^T)^{-1} \right) \mathbf{1}} $, as can be seen from the first order optimality conditions. Then, the overall optimal price is found by comparing the monopolist's profits achieved at the optimal solutions of the constrained subproblems. The complexity of the algorithm is $O(n^4)$, since there are at most $n$ such subproblems (again from Theorem  \ref{thm:singleprice}) and each such subproblem simply involves a matrix inversion ($O(n^3)$) in computing the centrality gain and the maximum achievable profit.



\subsection{The Case of Two Prices: Full and Discounted}
\label{subsec:two}

In this subsection, we assume that the monopolist can choose to offer the good in one of two prices, $p_L$ and $p_H
$ $(p_L
<p_H
)$ that are exogenously defined. For clarity of exposition we call $p_L
$ and $p_H$ the discounted and the full price respectively. The question that remains to be studied is to which agents should the monopolist offer the discounted price, so as to maximize her revenues. 
We state the following assumption that significantly simplifies the exposition.
\begin{assumption} \label{ass:exogenous}The exogenous prices $p_L, p_H$ are such that $p_L, p_H<\min_{i \in \cal{I}} a_i$.
\end{assumption}
Note that under Assumption \ref{ass:exogenous}, Equation \eqref{eq:br_basic} implies that all agents purchase a positive amount of the good at equilibrium, regardless of the actions of their peers. As shown previously, the vector of consumption levels satisfies
$
\mathbf{x}= \Lambda^{-1} (\mathbf{a}-\mathbf{p}+G\mathbf{x}),
$
and hence
$
\mathbf{x}= (\Lambda-G)^{-1} (\mathbf{a}-\mathbf{p}).
$
An instance of the monopolist's problem can now be written as:
\begin{equation}
\begin{aligned}	
\notag
(OPT) \quad \max & \quad (\mathbf{p}-c\mathbf{1})^T (\Lambda - G)^{-1} (\mathbf{a}-\mathbf{p})  \\
st. & \quad p_i \in \{p_L
,p_H
\} \quad \mbox{for all $i\in \cal{I}$}, 
\end{aligned}
\end{equation}
where $\Lambda\succ 0$ is a diagonal matrix, $G$ is such that $G\geq 0$, $diag(G)=0$ and Assumption~1 holds.

Let $p_N \triangleq \frac{p_H+p_L}{2}$, $\delta \triangleq p_H-p_N$, $\hat{\mathbf{a}} \triangleq {\mathbf{a}}-p_N$ and  $\hat{c} \triangleq p_N-c  \geq \delta $. Using these variables, and noting that any feasible price allocation  can be expressed as $\mathbf{p}=\delta \mathbf{y} +p_N$, where $y_i\in \{-1,1\}$,  OPT can alternatively be expressed as
\begin{equation}
\label{eq:optAlternative}
\begin{aligned}	
 \max & \quad ({\delta \mathbf{y} + \hat{c} {\mathbf{1}}})^T (\Lambda - G)^{-1} (\hat{\mathbf{a}}-\delta \mathbf{y} )  \\
s.t. & \quad y_i \in \{-1,1\} \quad \mbox{for all $i\in \cal{I}$}.
\end{aligned}
\end{equation}
We next show that OPT is NP-hard, and provide an algorithm that achieves an approximately optimal solution. To obtain our results, we relate the alternative formulation of OPT in \eqref{eq:optAlternative} to   the MAX-CUT problem (see~\cite{garey1979computers,Goemans:1995}).


\begin{theorem}
\label{thm:NP}
Let Assumptions \ref{ass:concavityCond}, \ref{ass:all_buy} and  \ref{ass:exogenous} hold. Then, the monopolist's optimal pricing problem, i.e., problem OPT, is NP-hard.
\end{theorem}
Finally, theorem \ref{theo:boundW_pricing} states that there exists an algorithm that provides a solution with a provable approximation guarantee.
\begin{theorem} \label{theo:boundW_pricing}
Let Assumptions  \ref{ass:concavityCond} and \ref{ass:exogenous} hold and $W_{OPT}$  denote the optimal profits for the monopolist, i.e., $W_{OPT}$ is the optimal value for problem OPT. Then, there exists a randomized polynomial time algorithm, that outputs a solution with objective value $W_{ALG}$ such that $E[W_{ALG}]+ m  >  0.878 (W_{OPT}+m),$ where
\[m =\delta^2 \mathbf{1}^T A \mathbf{1} + 
\delta \mathbf{1}^T  \left|  A  \hat{\mathbf{a}}  -   A^T \hat{c}  \mathbf{1}\right| -\hat{c} \mathbf{1}^T A \hat{\mathbf{a}} -2 \delta^2 Trace(A)
,\]  and $A=(\Lambda-G)^{-1}$
\end{theorem}
Clearly, if $m\leq 0$, which, for instance is the case when $\delta$ is small, this algorithm provides at least an $ 0.878$-optimal solution of the problem.

In the remainder of the section, we provide a characterization of the optimal  prices in OPT. 
In particular, we argue that the pricing problem faced by the monopolist is equivalent to finding the cut with maximum weight in an appropriately defined weighted graph. For simplicity, assume that $b_i=b_0$ for all $i$ and $\left( (\Lambda-G)^{-1}  \hat{\mathbf{a}}- \hat{c} (\Lambda-G)^{-T} \mathbf{1}   \right)=0$ (which holds, for instance when $\hat{\mathbf{a}}=\hat{c}{\mathbf{1}}$, or equivalently  $\mathbf{a}-p_N = (p_N-c) {\mathbf 1}$, and $G=G^T$).
Observe that in this case, the alternative formulation of the profit maximization problem in 
\eqref{eq:optAlternative}, can equivalently be written as (after adding a constant to the objective function, and scaling):
\begin{equation}
\label{eq:optAlternative3}
\begin{aligned}	
 \max & \quad \alpha- \mathbf{y}  (\Lambda - G)^{-1} \mathbf{y}   \\
s.t. & \quad y_i \in \{-1,1\} \quad \mbox{for all $i\in \cal{I}$},
\end{aligned}
\end{equation}
where $\alpha=\sum_{ij}(\Lambda - G)^{-1}_{ij}$. It can be seen that this optimization problem is equivalent to an instance of the MAX-CUT problem, where the cut weights are given by the off diagonal entries of $(\Lambda - G)^{-1}$ (see \cite{garey1979computers,Goemans:1995}).
 On the other hand observe that $(\Lambda - G)^{-1}\mathbf{1}=\frac{1}{2b_0} (I- \frac{1}{2b_0} G)^{-1}\mathbf{1}$,  hence,  the $i$th row sum of the entries of the matrix $(\Lambda - G)^{-1}$ is proportional to the centrality of the $i$th agent in the network. Consequently,  the $(i,j)$th entry of the matrix  $(\Lambda - G)^{-1}$, gives a measure of  how much the edge between $i$ and $j$ contributes to the centrality of agent $i$. Since the MAX-CUT interpretation suggests that the optimal solution of the pricing problem is achieved by maximizing the cut weight, it follows that the optimal solution of this problem price differentiates the agents who affect the centrality of each other significantly.
 
\section{How valuable is it to know the network structure?}
\label{sec:value}
Throughout our analysis, we have assumed that the monopolist has perfect knowledge of the interaction structure of her consumers and can use it optimally when choosing her pricing policy. In this section, we ask the following question: when is this information most valuable? In particular, we compare the profits generated in the following two extremes: (i) the monopolist prices optimally assuming that no network externalities are present, i.e., $g_{ij}=0$ for all $i,j\in {\cal I}$ (however, consumers take network externalities into account when deciding their consumption levels) (ii) the monopolist has perfect knowledge of how consumers influence each other, i.e., knows the adjacency matrix $G$, and can perfectly price discriminate (as in Subsection \ref{subsec:perfect}).
We will denote the profits generated in these settings by $\Pi_0$ and $\Pi_N$ respectively. The next lemma provides a closed form expression for                                                        $\Pi_0$ and $\Pi_N$.

\begin{lemma} \label{lem:profits}
Under Assumptions \ref{ass:concavityCond} and \ref{ass:all_buy}, the profits $\Pi_0$ and $\Pi_N$ are given by:
\begin{equation}
\Pi_0= \left( \frac{\mathbf{a}-c\mathbf{1}}{2} \right) ^T (\Lambda - G)^{-1}  \left( \frac{\mathbf{a}-c\mathbf{1}}{2} \right)
\end{equation}
and
\begin{equation}
\Pi_N= \left( \frac{\mathbf{a}-c\mathbf{1}}{2} \right) ^T 
\left(\Lambda - \frac{G+G^T}{2} \right)^{-1}  
\left( \frac{\mathbf{a}-c\mathbf{1}}{2} \right).
\end{equation}
\end{lemma}

The impact of network externalities in the profits is captured by the ratio $\frac{\Pi_0}{\Pi_N}$. 
For any problem instance,  with fixed parameters $\mathbf{a},c, \Lambda, G$ this ratio can be computed using Lemma \ref{lem:profits}. The rest of the section, focuses on relating this  ratio to the properties of the underlying network structure.
To simplify the analysis, we make the following assumption.

\begin{assumption} \label{ass:concavityCond2}
The matrix $\Lambda - G$ is positive definite.
\end{assumption}
Note that if $\Lambda - G$ is not symmetric,  we  still refer to this matrix as positive definite if $\mathbf{x}^T (\Lambda - G)\mathbf{x} >0$ for all $\mathbf{x}\neq 0$.
A sufficient condition for Assumption \ref{ass:concavityCond2} to hold can be given in terms of the diagonal dominance of  $\Lambda-G$. For instance, this assumption holds\footnote{This claim immediately follows from the Gershgorin circle theorem (see \cite{golub1996matrix}).}, if 
for all $i\in {\cal I}$, $b_i> \sum_{j\in {\cal I}} g_{ij}$ and
$b_i> \sum_{j\in {\cal I}} g_{ji}$. 

 Theorem \ref{theo:pricingGain} provides bounds on $\frac{\Pi_N}{\Pi_0}$ using the spectral properties of $\Lambda-G$.
\begin{theorem} \label{theo:pricingGain}
Under Assumptions \ref{ass:concavityCond}, \ref{ass:all_buy} and \ref{ass:concavityCond2}, 
\begin{equation}
0\leq
\frac{1}{2}+  \lambda_{min}\left(\frac{ MM^{-T}+M^TM^{-1}}{4} \right)
 \leq 
 \frac{\Pi_0}{\Pi_N}  
 \leq 
\frac{1}{2}+
  \lambda_{max}
  \left( \frac{ MM^{-T}+M^{T}M^{-1}}{4} \right)   \leq 1,
\end{equation}
where $M=\Lambda-G$ and $\lambda_{min}(\cdot)$, $\lambda_{max}(\cdot)$ denote the minimum and the maximum eigenvalues of their arguments respectively.
\end{theorem}

If the underlying network structure is symmetric, i.e., $G=G^T$, then $MM^{-T}=M^TM^{-1}=I$ and the bounds in Theorem \ref{theo:pricingGain} take the following form
\begin{equation}
\frac{1}{2}+  \lambda_{min}\left(\frac{ MM^{-T}+M^TM^{-1}}{4} \right)
= 
 \frac{\Pi_0}{\Pi_N}  
=
\frac{1}{2}+
  \lambda_{max}
  \left( \frac{ MM^{-T}+M^{T}M^{-1}}{4} \right)  = 1.
\end{equation}
This is consistent with Corollary \ref{cor:symmetricNetwork}, in which we show that if the network is symmetric then the monopolist does not gain anything by accounting for network effects. As already mentioned in the introduction, the benefit of accounting for network effects is proportional to how asymmetric the underlying interaction structure is. The minimum and maximum eigenvalues of matrix $\left( \frac{ MM^{-T}+M^{T}M^{-1}}{4} \right)$ that appear in the bounds of Theorem \ref{theo:pricingGain} quantify this formally, as they can be viewed as a measure of the deviation from symmetric networks. 

Finally, we provide a set of simulations, whose goal is twofold: first, we show that 
the bounds of Theorem \ref{theo:pricingGain} are quite tight by comparing them to the actual value of the ratio of profits (which can be directly computed by Lemma \ref{lem:profits}) and, second, we illustrate that accounting for network effects can significantly boost profits, i.e., that the ratio can be much lower than 1. 
In all our simulations we choose the parameters so that $M=\Lambda-G$ is a positive definite matrix.

\paragraph{Star Networks:} In our first set of simulations, we consider star networks with $n=100$ agents. In particular, there is a central agent (without loss of generality agent $1$), which has edges to the remaining agents, and these are the only edges in the network. Consider the following two extremes:
\begin{enumerate}
\item[(1)]  The central agent is influenced by all her neighbors but does not influence any of them, i.e., if we denote the corresponding interaction matrix by $G^1$, then $G^1_{ij}=1$ if $i=1$, $j\neq i$, and $G^1_{ij}=0$  otherwise.
\item[(2)]  The central agent influences all her neighbors but is not influenced by any of them, i.e., if we denote the corresponding interaction matrix by $G^2$, then $G^2_{ij}=1$ if $j=1$, $j\neq i$, and $G^2_{ij}=0$  otherwise.
\end{enumerate}
We compute the ratio of profits $\frac{\Pi_0}{\Pi_N}$ for a class of network structures given by matrices $G^\alpha= \alpha G^1 + (1-\alpha) G^2$, where $\alpha\in [0,1]$ ($\alpha=1$ and $\alpha=0$ correspond to the two extreme scenarios described above). 
In order to isolate the effect of the network structure, we assume that 
$a_i=a_1$, $b_i=b_1$ and $c_i=c$ for all $i\in {\cal I}$.
In particular,  in our first simulation we set $b_i= n/10$ and in the second simulation we set $b_i= n/20$  for all $i\in {\cal I}$.
For both simulations
 we set  $a_i-c=1$ for all $i\in {\cal I}$.

\begin{figure}[ht]
\centering
\subfloat{ \label{tab:matchingPenniesL}
\includegraphics*[width=8cm,height=6cm]{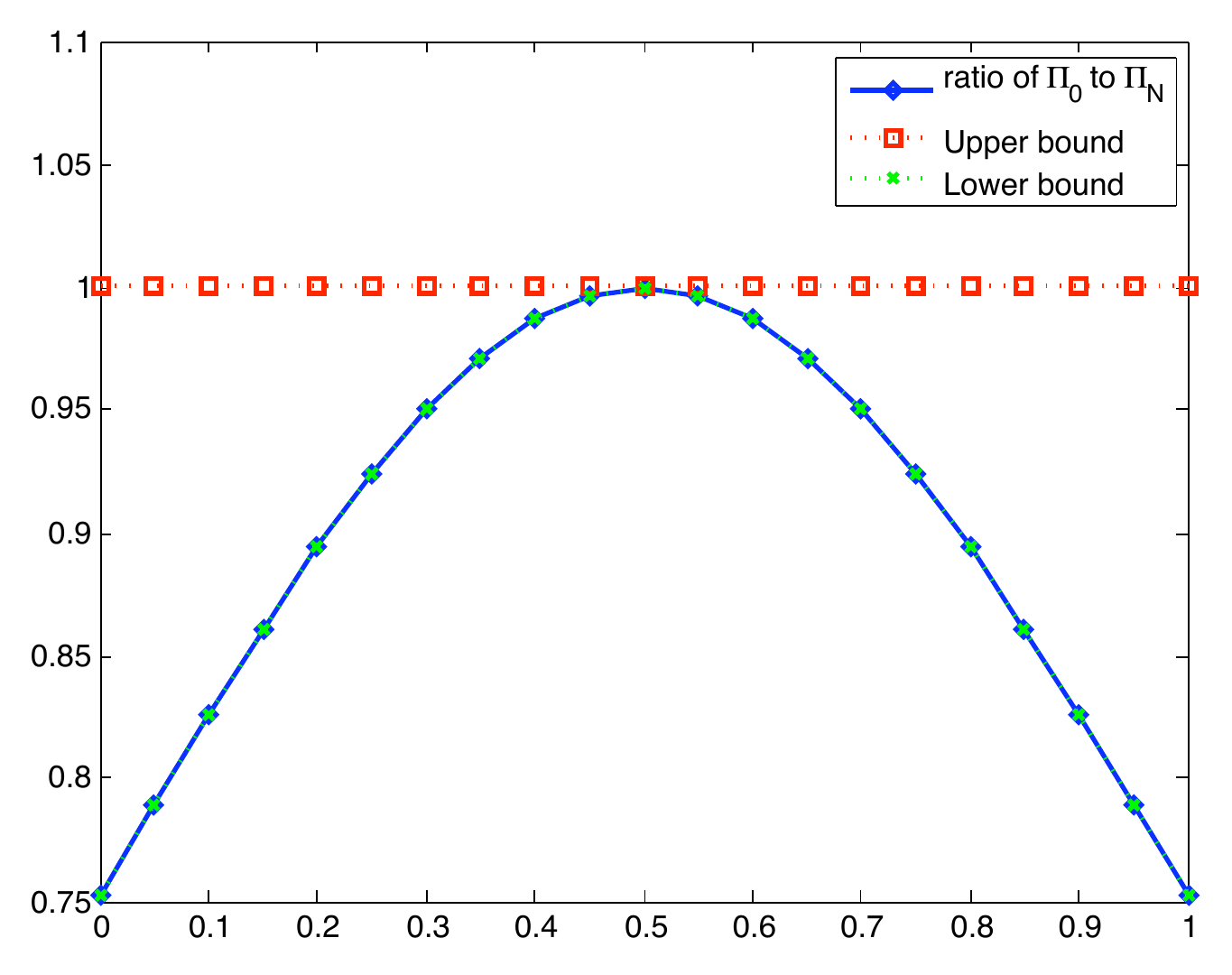}}
\subfloat{ \label{tab:matchingPenniesR}
\includegraphics*[width=8cm,height=6cm]{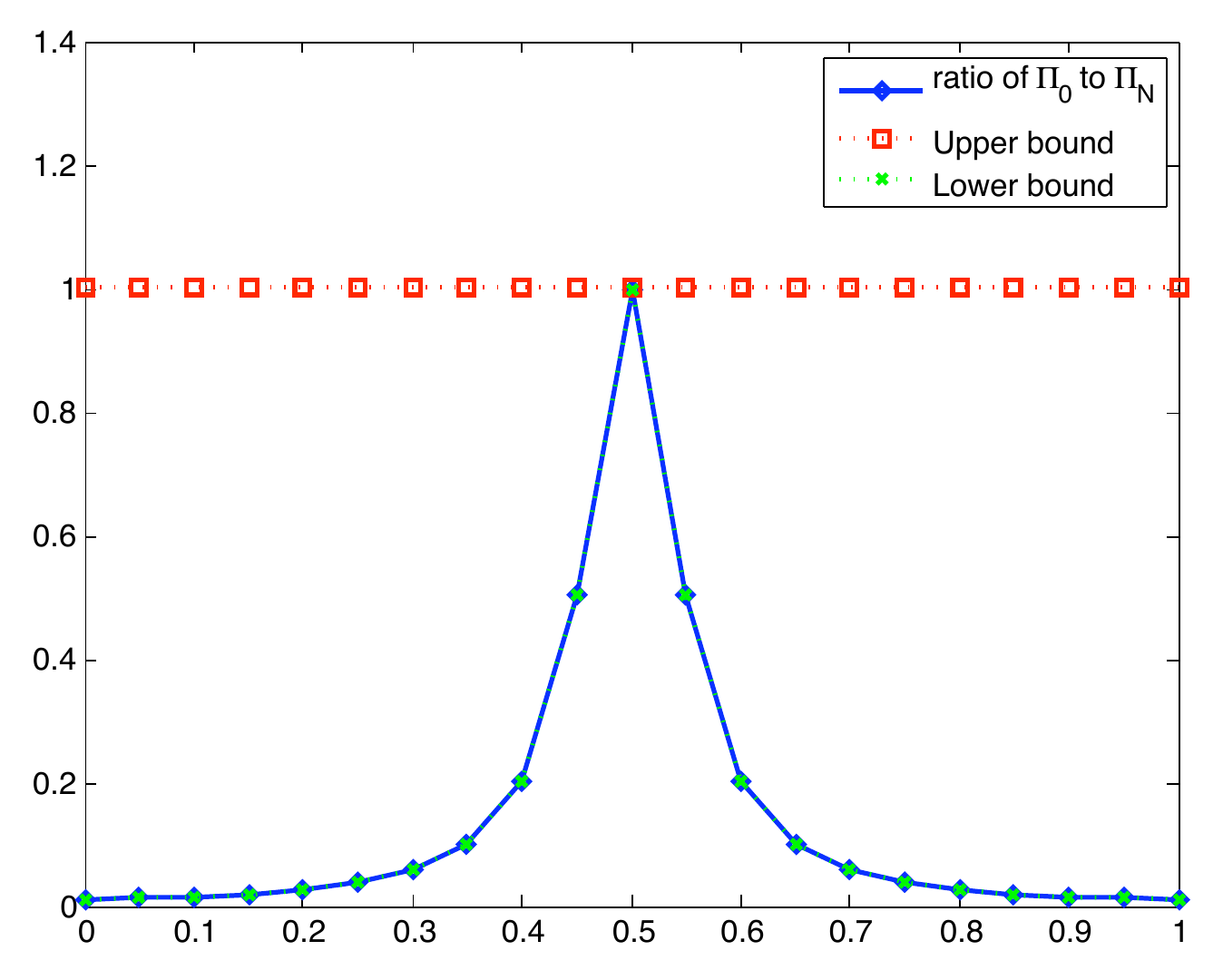}
}
\caption{Star Networks. Left: $b_i=n/10$, Right: $b_i=n/20$ for all $i\in {\cal I}$}        
\label{fig:star}
\end{figure}

The  results are presented in Figure \ref{fig:star}. In both simulations, the lower bound equals to the ratio $\frac{\Pi_0}{\Pi_N}$, implying that the bound provided in the theorem is tight. The upper bound seems to be equal to $1$ for all $\alpha$. 
 When $\alpha=\frac{1}{2}$, network effects become irrelevant, as the network is symmetric. On the other hand, for $\alpha=0$ and $\alpha=1$, i.e., when the star network is most ``asymmetric'', accounting for network effects leads to a $15 \% $ increase in profits when $b_i=n/10$ and to a $100-$fold increase when $b_i=n/20$. Choosing smaller $b_i$ increases the relative significance of network effects and, therefore, the increase in profits is much higher in the second case, when $b_i=n/20$. 
Although star networks are extreme, this example showcases that taking network effects into consideration can lead to significant improvements in profits.

\paragraph{From Asymmetric to Symmetric Networks :}
In this set of simulations, we replicate the above for arbitrary asymmetric networks. Again, we consider two extreme settings: let $U$ denote a fixed upper triangular matrix, and define the interaction matrices $G^1= U$ and $G^2= U^T$. The first, $G^1$, corresponds to the case, where agent $1$ is influenced by all her neighbors, but does not influence any other agent, and $G^2$ corresponds to the polar opposite where agent $1$ influences all her neighbors. As before, we plot the ratio of profits for a class of matrices parameterized by $\alpha\in[0,1]$, $G^\alpha= \alpha G^1 + (1-\alpha) G^2$. Specifically, we randomly generate $100$ upper triangular matrices $U$ (each none zero entry is an independent random variable, uniformly distributed in $[0,1]$).
We again consider two cases: $b_i=n/2$ and $b_i=n/3$ for all $i\in {\cal I}$.
 For each of these cases and randomly generated instances,  assuming   $a_i-c=1$ for all $i\in {\cal I}$, we obtain the ratio  $\frac{\Pi_0}{\Pi_N} $ and the bounds as given by Theorem \ref{theo:pricingGain}. The plots of the corresponding averages over all randomly generated instances are given in Figure \ref{fig:triangular}.

\begin{figure}[ht]
\centering
\subfloat{ \label{tab:matchingPenniesL}
{\includegraphics*[width=8cm,height=6cm]{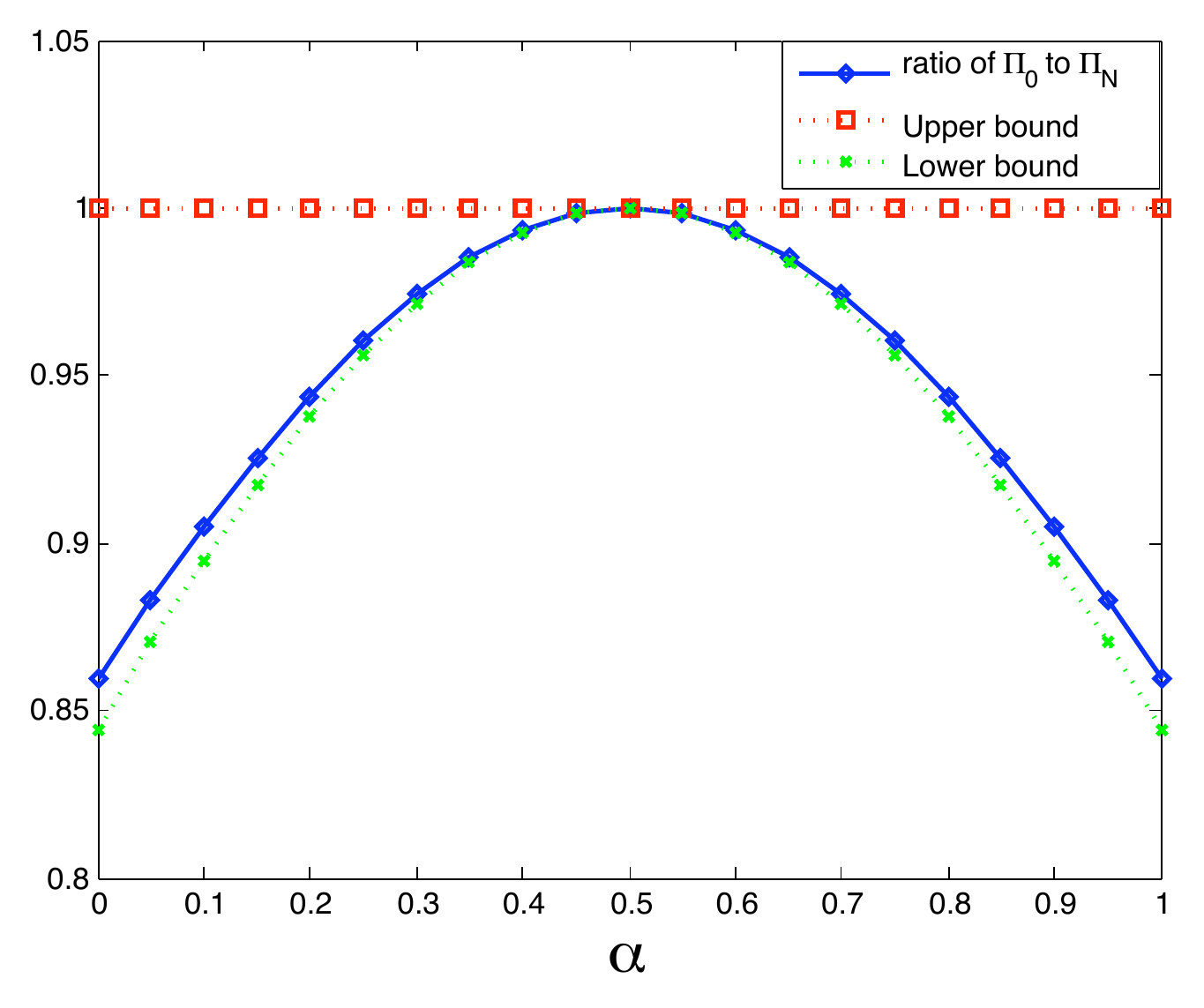}}}
\subfloat{ \label{tab:matchingPenniesR}
{\includegraphics*[width=8cm,height=6cm]{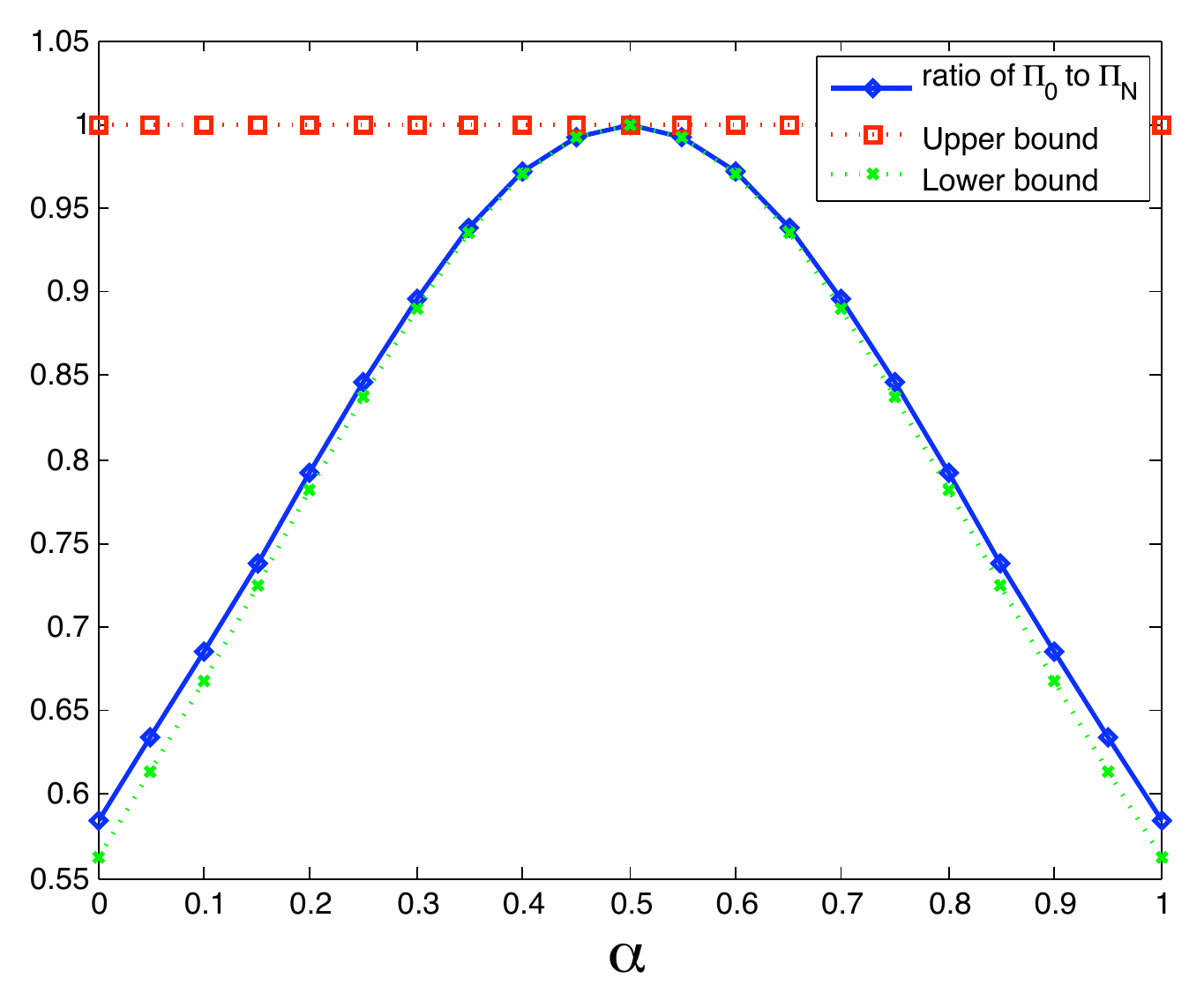}}}
\caption{Random asymmetric matrices.  Left: $b_i=n/2$, Right: $b_i=n/3$ for all $i\in {\cal I}$}  
\label{fig:triangular}
\end{figure}

Similar to the previous set of simulations, when $\alpha=\frac{1}{2}$, i.e., the network is symmetric, there is no gain in exploiting the network effects. On the other hand, for $\alpha=0$ and $\alpha=1$, i.e., when the network is at the asymmetric extremes, exploiting network effects can boost profits by almost $15 \%$ or  $40 \%$ depending on the value of $b_i$. Consistently with our earlier simulations, we observe that when $b_i$ is smaller, exploiting network effects leads to a more significant improvement in the profits. Note that for this network, the lower bound is not tight.
  
\paragraph{Preferential Attachment Graphs:} 
Finally, we consider networks that are generated according to a preferential attachment process, which is prevalent when modeling interactions in social networks. Networks are generated according to this process as follows: initially, the network consists of two agents and at each time instant a single agent is born and she is linked to two other agents (born before her) with probability proportional to their degrees. The process  terminates when the population of agents is $100$.

Given a random graph generated according to the process above, consider the following two extremes: (i) only newly born agents influence agents born earlier, i.e., the influence matrix $G^1$ is such that $G^1_{ij}>0$ for all $i,j$ that are linked in the preferential attachment graph, and $j$ is born after $i$, (ii) only older agents influence new agents, i.e., the influence matrix $G^2$ is such that $G^2_{ij}>0$ for all $i,j$ that are linked in the preferential attachment graph, and $j$ is born before $i$. We assume that the non-negative entries in each row of $G^1$ are equal and such that $G^1_{ij}=\frac{1}{d_i}$, where $d_i$ is the number of nonnegative entries in row $i$ (equal influence) and similarly for $G_2$. As before, we consider a family of networks parametrized by $\alpha$: $G^\alpha=\alpha G^1 + (1-\alpha) G^2$. The interaction matrix $G^\alpha$ models the situation, in which agents weigh the consumption of the agents that are ``born'' earlier by $1-\alpha$, and that of the new ones by $\alpha$. Note that since $G^1$ and $G^2$ are normalized separately, in this model $G^\alpha$ need not be symmetric, and in fact it turns out that for all $\alpha$ there is a profit loss by ignoring network effects.

In this model we consider two values for $b_i$: $b_i=2$ and $b_i=1.5$ for all $i\in {\cal I}$.
Also, we impose the symmetry conditions, $a_i-c=1$ for all $i\in {\cal I}$. 
Note that by construction, each preferential attachment graph is a random graph. For each $\alpha$, we generate $100$ graph instances and report the averages of $\frac{\Pi_0}{\Pi_N} $ and the bounds over all instances.

\begin{figure}[ht]
\centering
\subfloat{ \label{tab:matchingPenniesL}
{\includegraphics*[width=8cm,height=6cm]{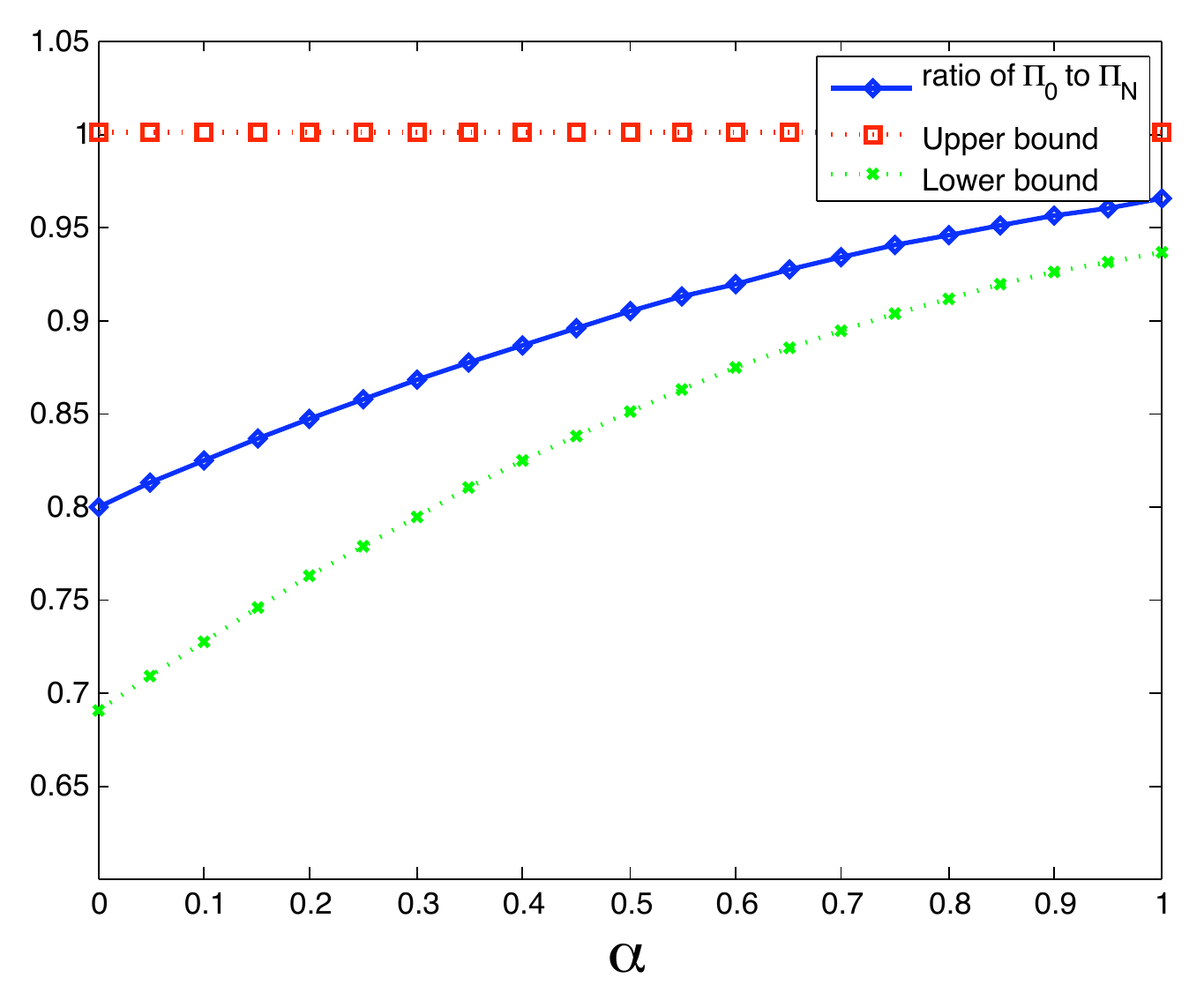}}}
\subfloat{ \label{tab:matchingPenniesR}
{\includegraphics*[width=8cm,height=6cm]{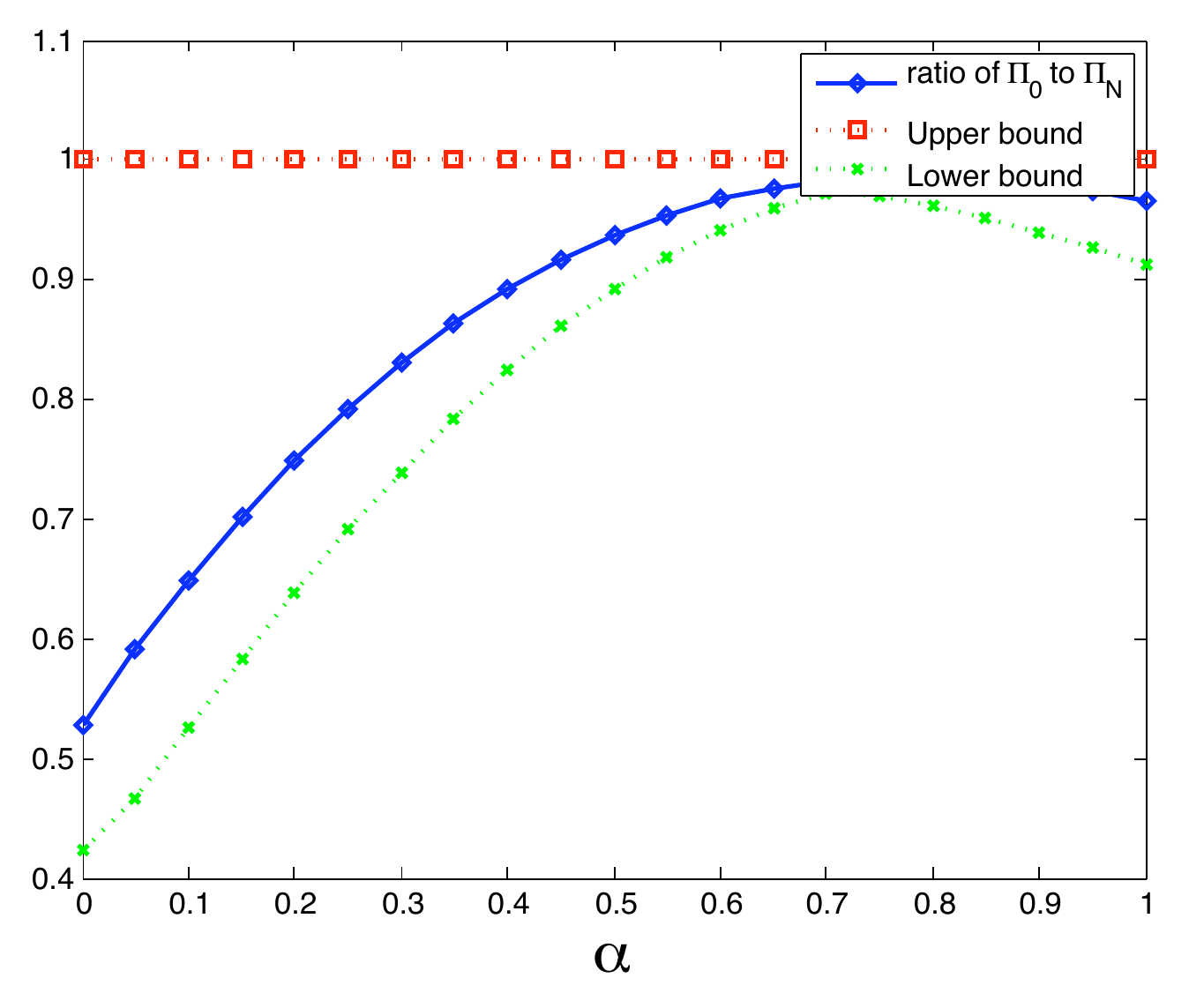}}}
\caption{Preferential attachment network example.  Left: $b_i=2$, Right: $b_i=1.5$ for all $i\in {\cal I}$}  
\label{fig:triangular}
\end{figure}

The plots are not symmetric, since as mentioned above $G^1$ and $G^2$ are normalized differently. Interestingly, the profit loss from ignoring network effects is larger when older agents influence agents born later ($\alpha=0$). This can be explained by the fact that older agents are expected to have higher centrality and act as interaction hubs for the network. As before, we see a larger improvement in profits when $b_i$ is small.

\section{Conclusions}
The paper studies a stylized model of pricing of divisible goods (services) over social networks, when consumers' actions are influenced by the choices of their peers. We provide a concrete characterization of the optimal scheme for a monopolist under different restrictions on the set of allowable pricing policies when consumers behave according to the unique Nash equilibrium profile of the corresponding game. We also illustrate the value of knowing the network structure by providing an explicit bound on the profit gains enjoyed by the monopolist due to this knowledge. 

Certain modeling choices, i.e., Assumptions \ref{ass:concavityCond}, \ref{ass:all_buy} and \ref{ass:exogenous}, were dictated by the need for tractability and were also essential for clearly illustrating our insights. For example, removing Assumption \ref{ass:all_buy} or \ref{ass:exogenous}  would potentially lead to a number of different subgame perfect equilibria of the two-stage pricing game faced by the monopolist. Although all these equilibria would share similar structural properties as the ones we describe (and would lead to the same profits for the monopolist), a clean characterization of the optimal prices (in closed form) would not be possible. Thus, we decided to sacrifice somewhat on generality in exchange to providing simple expressions for the optimal choices for the monopolist and clearly highlighting the connections with notions of centrality established in sociology. That being said, we expect that our analysis holds for more general environments.

Throughout the paper, we consider a setting of static pricing: the monopolist first sets prices and then the consumers choose their usage levels. Moreover, the game we define is essentially of complete information, since we assume that both the monopolist, as well as the consumers, know the network structure and the utility functions of the population. Extending our analysis by introducing incomplete information is an interesting direction for future research. Concretely, consider a monopolist that introduces a new product of unknown quality to a market. Agents benefit the monopolist in two ways when purchasing the product; directly by increasing her revenues, and indirectly by generating information about the product's quality and making it more attractive to the rest of the consumer pool. What is the optimal (dynamic) pricing strategy for the monopolist? 

Finally, note that in the current setup we consider a single seller (monopolist), so as to focus on explicitly characterizing the optimal prices as a function of the network structure. A natural departure from this model is studying a competitive environment. The simplest such setting would involve a small number of sellers offering a perfectly substitutable good to the market. Then, pricing may be even more aggressive than in the monopolistic environment: sellers may offer even larger discounts to ``central'' consumers, so as to subsequently exploit the effect of their decisions to the rest of the network. Potentially one could relate the \textit{intensity of competition} with the 
network structure. In particular, one would expect the competition to be less fierce when the network consists of disjoint large subnetworks, since then sellers would segment the market at equilibrium and exercise monopoly power in their respective segments.
\label{sec:conclusions}
\nocite{*} 
\bibliographystyle{aux_files/sty_files/ormsv080}
\bibliography{optimal_pricing}
\newpage

\section*{Appendix}

\subsection*{Proof of Theorem \ref{eq:uniqueEq}}

The proof makes use of the following lemmas.
\begin{lemma} \label{lem:invertible}
Under Assumption \ref{ass:concavityCond}, the spectral radius of $ \Lambda^{-1} G $ is smaller than $1$, and the matrix  $I- \Lambda^{-1} G$ is invertible.
\end{lemma}
\proof{Proof.}
 Let $v$ be an eigenvector of $\Lambda^{-1} G$ with $\lambda$ being the corresponding eigenvalue. Let $v_i$ be the largest entry of $v$ in absolute values, i.e., $|v_i|\geq |v_j|$ for all $j\in {\cal I}$. Since, $(\Lambda^{-1} G) v= \lambda v$, it follows that
\begin{equation*}
\notag
|\lambda v_i| = |(\Lambda^{-1} G)_i v| \leq  \sum_{j\in{ \cal I}} (\Lambda^{-1} G)_{ij} |v_j| \leq \frac{1}{2b_i} |v_i| \sum_{j\in {\cal I}}  g_{ij} < \frac{|v_i|}{2}
\end{equation*}
where $(\Lambda^{-1} G)_i$ denotes the $i$th row of $(\Lambda^{-1} G)$, the first and second inequalities use the fact that 
$ (\Lambda^{-1} G)_{ij}= \frac{g_{ij}}{2 b_i}\geq 0$,
 and
the last inequality follows from Assumption \ref{ass:concavityCond}. Since this is true for any eigenvalue-eigenvector pair, it follows that the spectral radius of $\Lambda^{-1} G$ is strictly smaller than $1$.

Note that each eigenvalue of $I- \Lambda^{-1} G$ can be written as $1-\lambda$ where $\lambda$ is an eigenvalue of $\Lambda^{-1} G$. Since the spectral radius of $\Lambda^{-1} G$ is strictly smaller than $1$ it follows that none of the eigenvalues of $I- \Lambda^{-1} G$ is zero, hence the matrix is invertible.
\Halmos\endproof

\begin{lemma}
Under Assumption \ref{ass:concavityCond}, the pure Nash equilibrium sets of games ${\cal G}=\{ {\cal I}, \{u_i\}_{{i\in {\cal I}}},$ ${[0, \infty)}_{i\in {\cal I}}  \}$ and $\bar{\cal G}=\{ {\cal I}, \{u_i\}_{i\in {\cal I}}, {[0, \bar{x}]}_{i\in {\cal I}}  \}$, where 
$\bar{x}= \max_i  \frac{|a_i-p_i|}{b_i}$, coincide.
\end{lemma}
\proof{Proof.}
The claim follows by proving that there is no equilibrium of game $\cal G$, such that $x_i>\bar{x}$ for some player $i$.  Assume for the sake of contradiction that such an equilibrium exists and let $i$ denote the agent with the largest consumption, $x_i$, at this equilibrium. Then, $x_i> \bar{x} \geq 0$ and
\begin{equation*}
\notag
x_i=\beta_i(\mathbf{x_{-i}})= \frac{a_i-p_i}{2 b_i}  + \frac{1}{2 b_i} \sum_{j\in {\cal I}} g_{ij} x_j \leq \frac{|a_i-p_i|}{2 b_i}  + \frac{1}{2 b_i} \sum_{j\in {\cal I}} g_{ij} x_i \leq \frac{|a_i-p_i|}{2 b_i}+ \frac{x_i}{2 },
\end{equation*}
where the last inequality follows from Assumption \ref{ass:concavityCond}. 
The above inequality implies that $x_i \leq \frac{|a_i-p_i|}{b_i}\leq \bar{x}$, which is a contradiction and, thus,  the claim follows.
\Halmos\endproof

We next show that $\bar{\cal G}$ is a supermodular game. Supermodular games are games that are characterized by strategic complementarities, i.e., the strategy sets of players are lattices, and  the marginal  utility of increasing a player's strategy raises with increases in the other players' strategies. For details and properties of these games, see \cite{topkis1998supermodularity}.

\begin{lemma}
The game $\bar{\cal G}=\{ {\cal I}, \{u_i\}_{i\in {\cal I}}, {[0, \bar{x}}]_{i\in {\cal I}}  \}$ is supermodular.
\end{lemma}
\proof{Proof.}
It is straightforward to see that the payoff functions are continuous, the strategy sets are compact subsets of $\mathbb{R}$, and for any players $i,j\in {\cal I}$, $\frac{\partial^2 u_i}{ \partial x_i\partial x_j } \geq 0$. Hence, the game is supermodular.
\Halmos\endproof

Now we are ready to complete the proof of the theorem. 
Since the set of equilibria of games ${\cal{G}}$ and $\cal{\bar{G}}$ coincide, we can focus on the equilibrium set of $\cal{\bar{G}}$. Since $\cal{\bar{G}}$ is a supermodular game, the equilibrium set has a minimum and a maximum element \cite{topkis1998supermodularity}. Let $\mathbf{x}$ denote the maximum of the equilibrium set and let set $S$ be such that $x_i>0$  only if $i\in S$. If $S=\emptyset$, there cannot be another equilibrium point, since $\mathbf{x}=0$ is the maximum of the equilibrium set. Thus, for the sake of contradiction, we assume that $S\neq \emptyset$ and there exists another equilibrium, $\mathbf{\hat{x}}$, of the game.

 By  supermodularity of the game, 
 it follows that
 $x_i \geq \hat{x}_i$ for all $i\in {\cal I}$.
Let $k\in\arg\max_{i\in {S}} x_i - \hat{x}_i$. Since $\mathbf{x}$ and $\mathbf{\hat{x}}$ are  not identical and $\mathbf{x}$ is the maximum of the equilibrium set, $x_k- \hat{x}_k>0$.

Note that at any equilibrium $\mathbf{z}$ of $\cal G$, no player has incentive to increase her consumption, thus $\left. \frac{\partial u_i(y_i,\mathbf{z_{-i}},p_i)}{\partial y_i }\right|_{y_i=z_i} \leq 0$. Moreover, if   $z_i>0$, since player $i$ does not have incentive to decrease its consumption, it also follows that $\left. \frac{\partial u_i(y_i,\mathbf{z_{-i}},p_i)}{\partial y_i }\right|_{y_i=z_i} = 0$.
Thus, from this condition and  \eqref{eq:quadUtilGen} it follows that
($G_k$ denotes the $k$th row of $G$) at equilibria $\mathbf{x}$ and $\mathbf{\hat{x}}$ we have
\begin{equation*}
\notag
\begin{aligned}
a_k-p_k&=2b_k x_k - G_k \mathbf{x} \\
a_k-p_k& \leq 2b_k \hat{x}_k - G_k \mathbf{\hat{x}},
\end{aligned}
\end{equation*}
where the latter condition holds with equality if $\hat{x}_k>0$.
Using these inequalities and Assumption \ref{ass:concavityCond}, it follows that
\begin{equation}
\notag
x_k - \hat{x}_k \leq \frac{1}{2b_k}  G_k (\mathbf{x} - \mathbf{\hat{x}})= \frac{1}{2b_k} \sum_j g_{kj }(x_j - \hat{x}_j) \leq \frac{x_k - \hat{x}_k}{2b_k} \sum_j g_{kj } <x_k - \hat{x}_k.
\end{equation}
We reach a contradiction, hence both $\cal G$ and $\bar{\cal G}$ have a unique equilibrium.

\subsection*{Proof of Lemma \ref{lem:monotoneDec}}
Consider a subset $S$ of the agents and consider the function $(\Lambda_S-G_S)^{-1}$$(\mathbf{a}_S
-\mathbf{p}_S)$. Observe that since the original network satisfies Assumption \ref{ass:concavityCond}, the network restricted to agents in $S$ also satisfies the same assumption.  By Lemma \ref{lem:invertible}, it follows that the matrix  $\Lambda_S-G_S$ is invertible and the spectral radius of $\Lambda_S^{-1} G_S$ is smaller than $1$. Therefore,
\begin{equation} \label{eq:expansionInv}
(\Lambda_S - G_S)^{-1}= (I - \Lambda_S^{-1} G_S)^{-1} \Lambda_S^{-1}= \sum_{k=0}^\infty (\Lambda_S^{-1} G_S)^k \Lambda_S^{-1},
\end{equation}
where the last equation follows since the spectral radius of $\Lambda_S^{-1} G_S$ is smaller than $1$. Observe that entries of $\Lambda_S^{-1} G_S$ and $\Lambda_S^{-1}$ are nonnegative. Thus it follows from \eqref{eq:expansionInv} that the entries of $(\Lambda_S - G_S)^{-1}$ are nonnegative. Therefore, each entry of the vector $  (\Lambda_S - G_S)^{-1} (\mathbf{a}_S
-\mathbf{p}_S)$ is weakly decreasing in $\mathbf{p}$.
Since this is true for any set $S$, by \eqref{eq:eq_characterization},  it follows that the equilibrium consumption $x_i(\mathbf{p})$ is weakly decreasing in $\mathbf{p}$ for all $i\in {\cal I}$.

\subsection*{Proof of Theorem \ref{theo:optPrice}}
The proof makes use of the following lemma, which states that under Assumptions \ref{ass:concavityCond} and \ref{ass:all_buy}, it is optimal for the monopolist to offer prices, so that all agents purchase a positive amount of the good.
\begin{lemma} \label{lem:postiveConsumption} 
Let Assumptions \ref{ass:concavityCond} and \ref{ass:all_buy} hold, and $\mathbf{p}^*$ denote an optimal solution of the first stage of the pricing-consumption game.
At the consumption equilibrium, $\mathbf{x}^*$, corresponding to $\mathbf{p}^*$,
all consumers purchase a positive amount of the good, i.e., $x^{*}_{i}>0$ for all $i \in \cal{I}$.
\label{lem:all_pos}
\end{lemma}
\proof{Proof.}For the sake of contradiction, let $({\mathbf{p}}^*,{\mathbf{x}}^*)$ be  such that $x_i^*=0$ for some $i \in \cal{I}$. We will construct a different price vector $\mathbf{p}^\prime$ by decreasing the price offered to player $i$ and increasing the prices offered to the rest of the agents. In our construction, we will ensure that if $\mathbf{p}^\prime$ is used, at equilibrium, agent $i$ purchases a positive amount of the good,  and the  consumptions of the remaining agents do not change. 
This will imply that the profit of the firm increases if $\mathbf{p}^\prime$ is used.

Consider agent $k$'s utility maximization problem. Recall that for a given price vector $\mathbf{p}$ the best response function satisfies:
\begin{equation}
\label{eq:best_res}
\beta_k(\mathbf{x_{-k}})=\max\left\{\frac{a_k-p_k}{2 b_k}  + \frac{1}{2 b_k} \sum_{j\in {\cal I}, j\neq k} g_{kj} x_j , 0 \right\}.
\end{equation}
Since at equilibrium $\mathbf{x}^*$, none of the agents have incentive to unilaterally deviate, it follows that 
${\mathbf{x}^*_{k}}=\frac{a_k-p_k^*}{2 b_k}  + \frac{1}{2 b_k} \sum_{j\in {\cal I}, j\neq k} g_{kj} x_j,$
if $x_k^*>0$, and ${\mathbf{x}^*_{k}}=0\geq  \frac{a_k-p_k^*}{2 b_k}  + \frac{1}{2 b_k} \sum_{j\in {\cal I}, j\neq k} g_{kj} x_j$
otherwise.

Consider a price vector $\mathbf{p^{\prime}}$ such that $p^{\prime}_i=c+\epsilon$, where $0<\epsilon<a_i-c$ (such an $\epsilon$ exists from Assumption \ref{ass:all_buy}) and  
\begin{equation}
\label{eq:new_price}
p^{\prime}_{j}=p_j^*
+ g_{ji} \left(\frac{a_i-p^{\prime}_i}{2 b_i}  + \frac{1}{2 b_i} \sum_{k\in {\cal I}, k\neq i} g_{ik} x_k^* \right), \textrm{ for all } j \neq i.
\end{equation}
Note that since $a_i>c+\epsilon=p_i^\prime$, it follows that $p^{\prime}_{j} > p_j^*$.
Let  $\{x^\prime_k\}$ be a consumption vector such that  $x^\prime_k = x_k^*$ if $k\neq i$ and $x^\prime_i= \frac{a_i-p^\prime_i}{2 b_i}  + \frac{1}{2 b_i} \sum_{j\in {\cal I}, j\neq i} g_{ij} x_j^* >0$.
It can be seen from \eqref{eq:best_res}, the solution of $\{x_k^*\}$ and \eqref{eq:new_price} that when prices are set to $\mathbf{p^\prime}$, each $x_k^\prime$ is the best response to $\mathbf{x}^\prime_{-k}$, and hence, the consumption vector, $\{x^\prime_k\}$, 
is the unique equilibrium point corresponding to $\mathbf{p}^\prime$.
Moreover, from \eqref{eq:best_res} we obtain that $x_i^{\prime}>\epsilon^{\prime}$ for some $\epsilon^\prime>0$.
 Since $p^{\prime}_i-c=\epsilon$, and $p^{\prime}_j \geq p_j$ for all $j \in \mathcal{I}, j\neq i$, it follows that 
 the monopolist increases her profits by at least $\epsilon \cdot \epsilon^{\prime}$ and the lemma follows.
\Halmos\endproof

\noindent Lemma \ref{lem:all_pos} and equation \eqref{eq:eqS_unique} imply that the optimal price $\mathbf{p}^*$ and the corresponding equilibrium vector $\mathbf{x}^*$ satisfy
\begin{equation}
\label{eq:eq1}
{\mathbf a}- \Lambda {\mathbf x}^* + G {\mathbf x}^* ={\mathbf p}^*.
\end{equation}
Thus, the problem that the monopolist is facing can be rewritten as:
\begin{displaymath}\begin{array}{ll} \max_{\mathbf{p}, \mathbf{x}} & \sum_{i} p_i x_i - c x_i \notag\\
\textrm{s.t.} & a_i - 2 b_i x_i +\sum_{j \in \mathcal{I}} g_{ij} x_j  - p_i = 0 , \textrm{ for every } i. \\
& x_i \geq 0,
\end{array}
\end{displaymath}
from which we obtain by the KKT conditions (and since we have already established that $x_i^*>0$ for all $i\in\cal{I}$):
\begin{equation}
\notag
{\mathbf a}-c{\mathbf 1}= \left( 2 \Lambda - (G+G^T)  \right){\mathbf x}^*,
\end{equation}
and hence
\begin{equation}
\notag
{\mathbf x}^*=\left(  \Lambda - \frac{G+G^T}{2}  \right)^{-1} \frac{{\mathbf a}-c{\mathbf 1}}{2}.
\end{equation}
Substituting ${\mathbf x}^*$ to \eqref{eq:eq1} the claim follows.

\subsection*{Proof of Theorem \ref{theo:optPriceSymPfirst}}
By Lemma \ref{lem:invertible}, $(\Lambda - G)$ is nonsingular, thus
 rearranging terms in  \eqref{eq:eq_optPrice}, it follows that
\begin{equation}
\label{eq:profitMInv}
\begin{aligned}
{\mathbf p} &= {\mathbf a} -(\Lambda - G) \left(\Lambda - G- \frac{G^T-G}{2}\right)^{-1} \frac{{\mathbf a}-c{\mathbf 1}
}{2} \\
&= {\mathbf a} - \left(I - \frac{G^T-G}{2} (\Lambda - G)^{-1}\right)^{-1} \frac{{\mathbf a}-c{\mathbf 1}
}{2}.
\end{aligned}
\end{equation}
To complete the proof, we need the matrix inversion lemma:
\begin{lemma}[Matrix inversion lemma]
Given square matrices of appropriate size,
\begin{equation}
\notag
(A-U D^{-1} V)^{-1}= A^{-1}+ A^{-1} U (D-VA^{-1}U)^{-1}VA^{-1}
\end{equation}
if $A$ and $D$ are nonsingular.
\end{lemma}
From this lemma, by setting $A, V=I$, $D=\Lambda - G$ and $U=\frac{G^T-G}{2}$, 
we obtain
\begin{equation*}
\begin{aligned}
\left(I - \frac{G^T-G}{2} (\Lambda - G)^{-1}\right)^{-1} & = I + \frac{G^T-G}{2} \left(\Lambda -G- \frac{G^T-G}{2}\right)^{-1} \\\notag & = I + \frac{G^T-G}{2} \left(\Lambda - \frac{G^T+G}{2} \right)^{-1}.
\end{aligned}
\end{equation*}
Thus, from \eqref{eq:profitMInv} it follows that
\begin{equation}
\label{eq:priceStructure}
\begin{aligned}
{\mathbf p} &= \frac{{\mathbf a}+c{\mathbf 1}
}{2} - \frac{G^T-G}{2}  \left(\Lambda - \frac{G^T+G}{2} \right)^{-1} \frac{{\mathbf a} - c{\mathbf 1}
}{2} .
\end{aligned}
\end{equation}
Using Assumption \ref{ass:symPlayers}, and substituting $\Lambda= {2b_0} I$ and $\mathbf{a}=a_0 \mathbf{1}$,  the vector of optimal prices can be rewritten as
\begin{equation}
\notag
\begin{aligned}
{\mathbf p} &= \frac{{a_0}+{c
}}{2} {\mathbf 1} + \frac{{a_0}-{c
}}{8b_0} G {\cal K} \left(\frac{G+G^T}{2}, \frac{1}{2b_0} \right)-\frac{{a_0}-{c
}}{8b_0} G^T {\cal K} \left(\frac{G+G^T}{2},\frac{1}{2b_0}\right).
\end{aligned}
\end{equation}

\subsection*{Proof of Theorem \ref{theo:optPriceSymP2}}
Immediate from \eqref{eq:priceStructure} and the definition of the weighted Bonacich centrality.

\subsection*{Proof of Lemma \ref{lem:monotoneDec2}}
The weakly decreasing  property of $\{x_i\}$ for each $i\in {\cal I}$, immediately follows from Lemma~\ref{lem:monotoneDec}.
The equilibrium characterization in \eqref{eq:eq_characterization} implies that if at price $p_0$ a set of agents $S$  consume a  positive amount of the good, then their consumption vector is given by 
\begin{equation} \label{eq:decreaseAllocation}
\mathbf{x}_S
(p_0)=   (\Lambda_S - G_S)^{-1} (\mathbf{a}_S-p_0 \mathbf{1}).
\end{equation}
As shown in the proof of Lemma \ref{lem:monotoneDec}, the entries of the matrix $(\Lambda_S - G_S)^{-1}$ are nonnegative. Since the matrix is invertible, none of the rows of the matrix $(\Lambda_S - G_S)^{-1} $ are identically equal to zero. Therefore, it follows that 
$ \left( (\Lambda_S - G_S)^{-1} \mathbf{1} \right)_i >0$, hence
\eqref{eq:decreaseAllocation} implies  that 
 $x_i$ is strictly decreasing in $p_0$ for all $i\in S$.

\subsection*{Proof of Theorem \ref{thm:singleprice}}

By Lemma \ref{lem:monotoneDec2}, it follows that the consumption vector at equilibrium is monotonically decreasing in $p$.  Moreover, if  the set of agents that purchase a positive amount of the good at equilibrium is given by $S$, then the consumption vector is as in \eqref{eq:decreaseAllocation}.
In order to prove the claim, we show  that among a set of agents $S$, who purchase a positive amount of the good,  the agent who first stops purchasing the good, as the price increases, is the one with the smallest centrality gain. Moreover, the price at which this agent stops is proportional to her centrality gain in the graph restricted to agent set $S$.

Consider a set of agents $S$ and price $p_0$ such that $p_0<a_i$ for all $i\in S$.  From  \eqref{eq:br_basic}, we obtain that all agents in $S$, have an incentive to purchase a positive amount of the good, regardless of the consumption levels of their peers. Thus, it follows that if this price is used, at equilibrium, all agents in $S$ purchase a positive amount of the  good. Using \eqref{eq:decreaseAllocation} and the definition of the weighted Bonacich centrality, the consumption vector can be rewritten as
\begin{equation} \label{eq:newCentrality}
\begin{aligned}
\mathbf{x}_S
(p_0) &=  \Lambda_S^{-1} (I - G_S \Lambda_S^{-1})^{-1} (\mathbf{a}_S-p_0 \mathbf{1})\\
&=
\Lambda_S^{-1} \tilde{K}(G_S, \Lambda_S^{-1},\mathbf{a}_S)- p_0 \Lambda_S^{-1} \tilde{ \cal K}(G_S, \Lambda_S^{-1},\mathbf{1}).
\end{aligned}
\end{equation}
Equivalently, for any $i\in S$, the consumption of player $i$ can be given as $$
x_i(p_0)=
\frac{1}{2b_i} \left( \tilde{\cal K}_i(G_S, \Lambda_S^{-1},\mathbf{a}_S)- p_0  \tilde{\cal K}_i(G_S, \Lambda_S^{-1},\mathbf{1}) \right).$$
Therefore, it follows that, when 
\begin{equation}
\label{eq:threshold}
p=\min_{i\in S}\frac{\tilde{\cal K}_i(G_S, \Lambda_S^{-1},\mathbf{a}_S)}{ \tilde{\cal K}_i(G_S, \Lambda_S^{-1},\mathbf{1})}=\min_{i\in S} H_i(G_S, \Lambda_S^{-1}, \mathbf{a}_S),
\end{equation}
then for the first time a group of  agents in $S$ stops purchasing the good.

It follows from \eqref{eq:threshold} that  if $p<p_1$ all agents in $\cal I$ purchase a positive amount of the good,
and  if the price is increased to $p_1=  \min_{i\in{\cal I}} H_i(G, \Lambda^{-1}, \mathbf{a})$, then agents in the set $D_1=\arg\min_{i\in{\cal I}} H_i(G, \Lambda^{-1}, \mathbf{a})$, stop purchasing the good. 
By monotonicity, these agents do not purchase the good when the price of the good is further increased. Furthermore, monotonicity also implies that the agents in the set ${\cal I}- D_1$ stop purchasing the good at a higher price. 
Using \eqref{eq:threshold} iteratively, it can be seen that  the agents in $D_k$ stop purchasing the good at price $p_k$ for $k\in \{1, \dots n\}$. 

Thus,
the first claim follows by construction of prices $p_k$ and the monotonicity of the consumption vector. 
The second claim follows from
 the fact that if $p<p_1$ then all agents purchase a positive amount of the good, and  if
 $p_{k}\leq p \leq p_{k+1}$, then only agents in ${\cal I}-\cup_{l=1}^k D_l=I_k$ purchase a positive amount of the good. 
Therefore, by \eqref{eq:eq_characterization}, the claim follows.

\subsection*{Proof of Theorem \ref{thm:NP}}

Recall that the MAX-CUT problem (with 0,1 weights) is defined as follows.
\begin{definition}[MAX-CUT Problem]
Let $G=(V,E)$ be an undirected graph and for all $i,j\in V$, define $g_{ij}$ such that $g_{ij}=1$ if $(i,j)\in E$ and $g_{ij}=0$ otherwise. Find the cut with maximum size, i.e., find a partition of the agent set $V$ into $S$ and $V-S$ such that the following sum is maximized:
\[\sum_{i\in S, j\in V-S} g_{ij}\] 
\end{definition}

Note that the MAX-CUT problem is equivalent to the following optimization problem:
\begin{equation*}
\notag
\begin{aligned}
\max & \quad \sum_{(i,j)\in E} W_{ij}  (1-x_i x_j)  \\
st. & \quad x_i \in \{-1,1\} \quad \mbox{for all $i\in V$},
\end{aligned}
\end{equation*}
where $W$ denotes the matrix of weights (we assume that $W_{ij}=W_{ji}\in \{0,1\}$). The optimal solution of the above problem corresponds to a cut as follows: let $S$ be the set of agents that were assigned value $1$ in the optimal solution. Then, it is straightforward to see that the value of the objective function corresponds to the size of the cut defined by $S$ and $V-S$.
We can further rewrite the optimization problem as:
\begin{equation*}
\notag
\begin{aligned}
 (P0)\quad  \min & \quad \mathbf{x}^T  W \mathbf{x}  \\
st. & \quad x_i \in \{-1,1\} \quad \mbox{for all $i\in V$}.
\end{aligned}
\end{equation*}
It is well-known that this problem is NP-hard \cite{garey1979computers}. Consider the following related problem:
\begin{equation*}
\begin{aligned}	
\notag
(P1)\quad \min & \quad \mathbf{x}^T  W \mathbf{x}  \\
st. & \quad x_i \in \{-1,1\} \quad \mbox{for all $i\in V$},
\end{aligned}
\end{equation*}
where $W$ is a symmetric matrix, with rational entries which satisfies ${0}<W^T=W<{1}$ (inequality is entrywise). We next show by reduction from MAX-CUT that P1 is also NP-hard.
\begin{lemma}
P1 is NP-hard.
\end{lemma}
\proof{Proof.}
We prove the claim by reduction from P0. Let $W$ be the weight matrix in an instance of P0. Then, let $W_\epsilon=\frac{1}{2}( \epsilon+W)$, where $\epsilon$ is a rational number such that $0<\epsilon< \frac{1}{2n^2}$ and $|V|=n$. Observe that for any feasible $\mathbf{x}$ in P0 or P1 it follows that
\begin{equation*}
2 \mathbf{x}^T W_\epsilon \mathbf{x} - n^2 \epsilon \leq \mathbf{x}^T W \mathbf{x} \leq 2 \mathbf{x}^T W_\epsilon \mathbf{x} + n^2 \epsilon.
\end{equation*}
Since the objective value of P0 is always an integer, and $n^2 \epsilon<\frac{1}{2}$, it follows that the cost of P0 for any feasible vector $\mathbf{x}$ can be obtained from the cost of P1 (with $W_\epsilon$) by scaling and rounding. Hence, it can be seen that from the optimal solution of the latter, we can immediately obtain the optimal solution and the value of the former (as rounding is a monotone operation). Therefore, since P0 is NP-hard it follows that P1 is also  NP-hard and the claim follows.
\Halmos\endproof

Next we prove Theorem \ref{thm:NP}, by using a reduction from P1 to OPT. We consider the special instances of OPT for which we have $G=G^T$, ${c}=0$, $\mathbf{a}=[a
, \cdots, a],
$ where $a=p_L+p_H$.
Observe that under this setting, Assumptions \ref{ass:all_buy} and \ref{ass:exogenous} hold and $\hat{\mathbf a}=p_N \mathbf{1}=\hat{c} {\mathbf{1}}$. Hence using \eqref{eq:optAlternative},  the instances of OPT can be rewritten as 
(by adding a constant and scaling the objective function)
\begin{equation*}
\begin{aligned}	
(F) \quad \min & \quad \mathbf{x}^T (\Lambda - G)^{-1}\mathbf{x} \\
s.t. & \quad x_i \in \{-1,1\} \quad \mbox{for all $i\in \cal{I}$}.
\end{aligned}
\end{equation*}
 
Next we show that any instance of P1 can be transformed into an instance of F (or equivalently OPT) where $\Lambda$ and $G$ are matrices with rational entries, $G=G^T\geq 0$ (the inequality is entry wise), $diag(G)=0$, $\Lambda_{ij}=0$ if $i\neq j$, $\Lambda_{kk}>0$,  $\Lambda$ and $G$ matrices satisfy Assumption~1.
Note that Assumption~1 is equivalent to requiring $(\Lambda - G)_k \mathbf{e}^k> 0$ for all $k$, where 
  $\mathbf{e}^k$ denotes the vector $k$th entry of which is equal to $\frac{1}{2}$, and the remaining entries are one, and   $(\Lambda - G)_k$ denotes the $k$th row of $(\Lambda - G)$.
Observe that these requirements on $\Lambda$ and $G$ ensure that the corresponding instance of OPT satisfies the assumptions of the theorem.

Consider an instance of P1 with $W>0$. Note that since $x_i^2=1$, P1 is equivalent to 
\begin{equation} \label{eq:optNew}
\begin{aligned}	
\quad \min & \quad \mathbf{x}^T  (W + \gamma I)  \mathbf{x}  \\
st. & \quad x_i \in \{-1,1\} \quad \mbox{for all $i\in V$},
\end{aligned}
\end{equation}
where we choose $\gamma$ as an integer such that $\gamma> 4 \max \left\{\rho(W), \frac{\sum_{i,j}W_{ij}}{\min_{ij} W_{ij}}\right\}$, and $\rho(\cdot)$ denotes the spectral radius of its argument.
Next, we show that this optimization problem is equivalent to an instance of F, by showing that $(W + \gamma I) = (\Lambda - G)^{-1}$, for some $\Lambda$ and $G$ satisfying the requirements above.

The definition of $\gamma$ implies that the spectral radius of $\frac{W}{\gamma}$ is smaller than $1$. 
 Therefore, it follows that
\begin{equation*}
(W+\gamma I)^{-1}=\frac{1}{\gamma} \left( I - \frac{W}{\gamma}+ \frac{W^2}{\gamma^2} \dots  \right)
\end{equation*}
Observe that for all $i,j \in \{1, 2 \dots n\}$,
\begin{equation*}
\begin{aligned}
\left( W- \frac{W^2}{\gamma} \right)_{ij}&= W_{ij} - \frac{ \sum_{k}W_{ik} W_{kj}}{\gamma} \geq W_{ij} - \frac{ \sum_{k}W_{ik}}{\gamma} \\
& \geq W_{ij}-\frac{\min_{kl} W_{kl}}{4} \geq  W_{ij}-\frac{ W_{ij}}{4}>0,
\end{aligned}
\end{equation*}
where the first inequality follows from the fact that $ {0}<W<{1}$ and the second inequality follows from the definition of $\gamma$.
Thus all entries of $\left( W- \frac{W^2}{\gamma} \right)$ are positive.
Rewriting $(W+\gamma I)^{-1}$ as 
\begin{equation*}
(W+\gamma I)^{-1}=\frac{1}{\gamma} \left( I - \frac{1}{\gamma} \left( {W}- \frac{W^2}{\gamma}  \right) - \frac{W^2}{\gamma^3} \left( {W}- \frac{W^2}{\gamma}  \right) \dots  \right).
\end{equation*}
and noting that all entries of $W$ and $\left( W- \frac{W^2}{\gamma} \right)$ are positive, the above equality implies that
the off diagonal entries of $(W+\gamma I)^{-1}$ are negative.
Thus, $(W+\gamma I)^{-1}= (\Lambda -G)$ for some diagonal matrix $\Lambda$ and for some $G\geq0$ with $diag(G)=0$. Moreover, since $W=W^T$, $G$ is also a symmetric matrix. 
Note that since the spectral radius of $\frac{W}{\gamma}$ is smaller than $1$, it also follows that
\begin{equation*}
\begin{aligned}
( (\Lambda -G)\mathbf{e}^k)_k&=((W+\gamma I)^{-1} \mathbf{e}^k)_k= \left( \frac{1}{\gamma} \left( I - \frac{W}{\gamma}+ \frac{W^2}{\gamma^2} \dots  \right) \mathbf{e}^k \right)_k.
\end{aligned}
\end{equation*}
Since $W>0$, it can be seen that $W^l>0$ for all $l\in \mathbb{Z}_+$.  Using this observation and the inequality $\mathbf{1} \geq \mathbf{e}^k > 0$ (entrywise), we obtain
\begin{equation*}
\begin{aligned}
( (\Lambda -G)\mathbf{e}^k)_k & \geq \frac{1}{\gamma} \left( \frac{1}{2} + \left(- \frac{W\mathbf{e}^k}{\gamma}- \frac{W^2 \mathbf{e}^k}{\gamma^2} \dots  \right)_k  \right)    \\
&\geq  \frac{1}{\gamma} \left( \frac{1}{2} + \left(- \frac{W\mathbf{1}}{\gamma}- \frac{W^2 \mathbf{1}}{\gamma^2} \dots  \right)_k  \right).
\end{aligned}
\end{equation*}
By the definition of $\gamma$, it follows that
$ \frac{ {W \mathbf{1}}}{\gamma} \leq \frac{ {\sum_{ij} W_{ij}} }{\gamma}\mathbf{1} \leq  \frac{\mathbf{1}}{4} $.
Therefore, the above inequality implies that
\begin{equation*}
\begin{aligned}
( (\Lambda -G)\mathbf{e}^k)_k&
\geq  \frac{1}{\gamma} \left( \frac{1}{2} -  \frac{{1}}{4 } \left(\sum_{l=0}^\infty \left( \frac{{1}}{4} \right)^l \right) \right)  = \frac{1}{\gamma} \left(\frac{1}{6} \right)>0.
\end{aligned}
\end{equation*}
Thus, Assumption \ref{ass:concavityCond} holds for the game defined with the matrices $\Lambda$ and  $G$.  Note that since the off diagonal entries of $\Lambda - G$ are nonpositive,  Assumption \ref{ass:concavityCond} implies that  the diagonal entries of $\Lambda$ are positive.

Therefore, problem P1 can be  reduced to an instance of F, by defining $\Lambda$ and $G$ according to $(W+\gamma I)^{-1}= (\Lambda -G)$. Thus, it follows that F and hence OPT are NP-hard.

\subsection*{Proof of Theorem \ref{theo:boundW_pricing}}
First, we describe a semidefinite programming (SDP) relaxation for the following optimization problem: 
\begin{equation} \label{eq:MC_form0}
\begin{aligned}
\max & \quad  \frac{1}{4} \sum_{i,j} w_{ij} (1-x_i x_j) \\
st. & \quad x_i \in \{-1,1\} \qquad \mbox{for all $i\in V$}.
\end{aligned}
\end{equation}
Note that  \eqref{eq:MC_form0} can be relaxed to
\begin{equation} \label{eq:MC_app0}
\begin{aligned}
\max & \quad  \frac{1}{4} \sum_{i,j} w_{ij} (1-\mathbf{\nu}_i \cdot \mathbf{\nu}_j) \\
st. & \quad \mathbf{\nu}_i \in S_n \qquad \mbox{for all $i\in V$} 
\end{aligned}
\end{equation}
where 
$\mathbf{\nu}_i \cdot \mathbf{\nu}_j$, denotes the regular inner product of vectors $\mathbf{\nu}_i, \mathbf{\nu}_j \in \mathbb{R}^n$, and
$S_n$ denotes the $n$-dimensional unit sphere, i.e., $S_n=\{\mathbf{x}\in \mathbb{R}^n| \mathbf{x} \cdot\mathbf{x}=1\}$.
We next show that \eqref{eq:MC_app0} leads to a semidefinite program. 

Consider a collection of vectors $\{\mathbf{\nu}_1, \cdots , \mathbf{\nu}_n\}$ such that $\mathbf{\nu}_i\in S_n$. Define
 a symmetric matrix $Y\in R^{n\times n}$, such that $Y_{ij}=\mathbf{\nu}_i \cdot \mathbf{\nu}_j$ and $Y_{ii}=1$. It can be seen that $Y=F^T F$, where $F\in \mathbb{R}^{n\times n}$ is such that $F=[\mathbf{\nu}_1, \mathbf{\nu}_2 \dots \mathbf{\nu}_n]$. This implies that $Y\succeq 0$. Conversely,
consider a positive semidefinite matrix $Y \in \mathbb{R}^{n\times n}$, such that $Y_{ii}=1$. 
Since $Y$ is positive semidefinite, there exists $F\in \mathbb{R}^{n\times n}$ (which can be obtained through the Cholesky factorization of the original matrix) such that $Y= F^T F$. 
Denote the columns of $F$ by $\mathbf{\nu}_i$, i.e., $F=[\mathbf{\nu}_1, \mathbf{\nu}_2 \dots \mathbf{\nu}_n]$. Since $Y_{ii}=1$, and $Y=F^TF$, it follows that $\mathbf{\nu}_i \cdot \mathbf{\nu}_i=1$. 
These arguments imply that the feasible set in \eqref{eq:MC_app0}, can equivalently be defined in terms of positive semidefinite matrices.
Hence, it follows that the optimization problem in \eqref{eq:MC_app0}, can be equivalently written as
\begin{equation} \label{eq:MC_app1}
\begin{aligned}
\max & \quad  \frac{1}{4} \sum_{i,j} w_{ij} (1-Y_{ij}) \\
st. & \quad Y_{i,i}=1  \qquad \mbox{for all $i\in V$} \\
& \quad Y\succeq 0.
\end{aligned}
\end{equation}

Next, we show how to obtain a provable approximation guarantee for binary quadratic optimization problems of the form:
\begin{equation}\label{eq:QBI1}
\begin{aligned}
\max& \quad \mathbf{x}^T Q \mathbf{x}+2\mathbf{d}^T \mathbf{x}+z  \\
s.t. & \quad x_i\in \{-1,1\}, \qquad \mbox{$i\in \{1,\dots , n\}$},
\end{aligned}
\end{equation}
where $Q$ and $\mathbf{d}$, $z$ have rational entries, i.e., $ Q\in \mathbb{Q}^{n\times n}$, $\mathbf{d}\in \mathbb{Q}^n$, and $z\in \mathbb{Q}$.
Observe that $\mathbf{x}^T Q \mathbf{x}=Trace({Q})+\mathbf{x}^T\tilde{Q} \mathbf{x}$, where $\tilde{Q}=Q-diag(Q)$ and $ x_i\in \{-1,1\}$. Thus, the diagonal entries of the $Q$ matrix can be expressed as a part of the constant term, and thus,  we can assume that   $diag(Q)=0$ without any loss of generality. 
Also, again without loss of generality, we can assume that the matrix $Q$ is symmetric, since 
$\mathbf{x}^T Q \mathbf{x}=\mathbf{x}^T Q^T \mathbf{x}= \mathbf{x}^T \frac{Q + Q^T}{2}\mathbf{x}$.

Consider the following optimization problem
\begin{equation} \label{eq:QBI2}
\begin{aligned}
\max& \quad [\mathbf{x}; y] ^T \hat{Q} [\mathbf{x};y]+{z}  \\
s.t. & \quad x_i\in \{-1,1\}, \qquad \mbox{$i\in \{1,\dots , n\}$},\\
& \quad y\in \{-1,1\},
\end{aligned}
\end{equation}
where 
\begin{equation} 
\hat{Q}= 
\left[ 
\begin{array}{c | c} 
Q & \mathbf{d} \\ \hline 
\mathbf{d}^T & 0
\end{array}\right].
\end{equation}
Note that given a feasible solution $[\mathbf{x};y]$ of \eqref{eq:QBI2}, another feasible solution  with the same objective value is $-[\mathbf{x};y]$. Therefore, given an optimal solution of \eqref{eq:QBI2}, another optimal solution where $y=1$, can be obtained. 
Since by construction $ [\mathbf{x}; y] ^T \hat{Q} [\mathbf{x};y] = \mathbf{x}^T Q \mathbf{x} + 2 y  \mathbf{d}^T \mathbf{x}$, it follows that the optimal $\mathbf{x}$ solution for \eqref{eq:QBI2} is also optimal for \eqref{eq:QBI1} and the optimal objective values for the two problems  are equal. Therefore, instead of solving \eqref{eq:QBI1}, we focus on \eqref{eq:QBI2}.

Following \eqref{eq:MC_form0}, and \eqref{eq:MC_app0}, we can relax \eqref{eq:QBI2} to:
\begin{equation} \label{eq:rel_QBI2}
\begin{aligned}
\max& \quad \sum_{ij} \mathbf{\nu}_i \cdot \mathbf{\nu}_j \hat{Q}_{ij} +z \\
s.t. & \quad \mathbf{\nu}_i\in S_{n+1}, \qquad \mbox{$i\in \{1,\dots, n, n+1\}$},
\end{aligned}
\end{equation}
and obtain an equivalent SDP (by defining $Y_{ij}= \mathbf{\nu}_i \cdot \mathbf{\nu}_j$) as follows:
\begin{equation} \label{eq:sdp_QBI2}
\begin{aligned}
\max& \quad \sum_{ij} Y_{ij} \hat{Q}_{ij} +z  \\
s.t. & \quad  Y_{ii}=1 \qquad \mbox{$i\in \{1,\dots, n, n+1\}$},\\
&\quad Y\succeq 0.
\end{aligned}
\end{equation}
Using this SDP relaxation, Algorithm \ref{alg2} provides an approximate solution to the original problem. We prove this, using a similar approach to \cite{Goemans:1995}.
\begin{algorithm}
\caption{: Compute  $\{x_1, \dots, x_n\}$, which is an approximate solution of \eqref{eq:QBI1} }
\label{alg2}
\begin{algorithmic}
\label{alg:MC}
\item[]
\item[] \textbf{STEP 1.}   Solve the SDP relaxation in \eqref{eq:sdp_QBI2}, find an optimal $Y$.
\item[] \textbf{STEP 2.}  Obtain the Cholesky factorization of $Y$, i.e., find $F$ such that $Y=F^TF$. Denote the $i$th column of $F$ by $\mathbf{\nu}_i$. Denote by $\nu_{n+1}$ the vector corresponding to the variable $y$ in \eqref{eq:QBI2}. 
\item[] \textbf{STEP 3.}  Let $r$ be a vector uniformly distributed on the unit sphere $S_{n+1}$.
\item[] \textbf{STEP 4.}  Let $S=\{i| r \cdot \mathbf{\nu}_i \geq 0\}$. If $n+1 \in S$, then set $x_i=1$ for all $i\in S\cap \{1,\dots,n\}$ and set the remaining $x_i$ to $-1$. Else if $n+1 \notin S$, then set $x_i=-1$ for all $i\in S\cap \{1,\dots,n\}$ and set the remaining $x_i$ to $-1$.
\item[] \textbf{Output:} $\{x_1,\dots, x_n\}$.
\end{algorithmic}
\end{algorithm}

%
\begin{proposition} \label{theo:QTHM}
Let $z\geq \sum_{i,j} |\hat{Q}_{ij}|$. Then, a solution given by Algorithm \ref{alg2}, achieves at least  $0.878$  times the optimal objective value of the original problem in \eqref{eq:QBI1}.
\end{proposition}
\proof{Proof.}
Let $W$ denote the objective value of  a solution the  algorithm provides, $W_M$ denote the optimal solution of the underlying quadratic optimization problem \eqref{eq:QBI1},  and $W_P$ denote the optimal value of the  SDP relaxation.
Let $\{\mathbf{\nu}_i\}$ denote the solution of SDP relaxation, then the corresponding  optimal  value can be given as
$$
W_P= \sum_{i,j}  {\hat{Q}_{ij} \mathbf{\nu}_i \cdot \nu _j}+z.
$$

It can be seen that
for solutions the algorithm provides, the probability\footnote{We assume that the range of the $\arccos$ function is $[0,\pi]$.} that agents $i$ and $j$ have opposite signs is 
$\frac{\arccos (\mathbf{\nu}_i \cdot \mathbf{\nu}_j)}{\pi}$, and similarly the probability that agents have the same sign is $1-\frac{\arccos (\mathbf{\nu}_i \cdot \mathbf{\nu}_j)}{\pi}$ (see \cite{Goemans:1995}). Thus, the expected contribution of these pair of agents to the objective function   is given by $\hat{Q}_{ij} \left(1-2\frac{\arccos (\mathbf{\nu}_i \cdot \mathbf{\nu}_j)}{\pi}\right)$. 
Hence, it follows that
 the expected value of a solution the algorithm provides   is given by
$$E[W]=  \sum_{i,j} \left(1-2\frac{arccos (\mathbf{\nu}_i \cdot \mathbf{\nu}_j)}{\pi}\right) \hat{Q}_{i,j} +z.$$
Since   $z\geq \sum_{i,j} |\hat{Q}_{ij}|$, it follows that both $W_M$ and $E[W]$ are nonnegative, also since $W_P$ corresponds to the optimal solution of the relaxation, it follows that $W_P\geq W_M$ . Using these it follows that 
$$
W_P= \sum_{i,j; \hat{Q}_{ij}>0}  {\hat{Q}_{ij} (1+\mathbf{\nu}_i \cdot \nu _j)}+ \sum_{i,j; \hat{Q}_{ij}<0}  {|\hat{Q}_{ij}| (1- \mathbf{\nu}_i \cdot \nu _j)}+z_2
$$
and 
$$E[W]=  \sum_{i,j; \hat{Q}_{ij}>0} \hat{Q}_{ij}\left(2-2\frac{arccos (\mathbf{\nu}_i \cdot \mathbf{\nu}_j)}{\pi}\right) + \sum_{i,j; \hat{Q}_{ij}<0} |\hat{Q}_{i,j}| 2\frac{arccos (\mathbf{\nu}_i \cdot \mathbf{\nu}_j)}{\pi} +z_2,$$
where $z_2=z- \sum_{i,j} |\hat{Q}_{ij}| \geq 0$. Since  the $\arccos$ function satisfies
$\frac{\arccos x}{\pi} \geq \frac{\alpha}{2} (1-x)$ and $1-\frac{\arccos x}{\pi} \geq \frac{\alpha}{2 }(1+x) $ for all $x\in[-1,1]$, where $\alpha \approx 0.878$ (see \cite{Goemans:1995}), it follows that $E[W]> 0.878 W_P \geq 0.878 W_M$.
\Halmos\endproof

\noindent This result can be extended, by relaxing the condition $z\geq \sum_{i,j} |Q_{ij}|$. To see this, we first  add a positive constant to the objective function of the original problem, ensuring that the modified problem satisfies this condition,  and then provide an approximation to this new problem. Note that the constant change in the objective function does not affect the output of the algorithm. The following corollary summarizes this result.
\begin{corollary} \label{cor:boundOnQuadGen}
Let $W$ denote the objective value of  a solution output by Algorithm \ref{alg2} and $W_M$ denote the optimal solution of the underlying quadratic optimization problem. Then,
$E[W]+\sum_{ij} |\hat{Q}_{ij}|-z  >  0.878 (W_M+\sum_{ij} |\hat{Q}_{ij}|-z)$.
\end{corollary}

Finally, when Assumption  \ref{ass:exogenous} holds,  using \eqref{eq:optAlternative}, and $A=(\Lambda-G)^{-1}$, the pricing problem of the firm can be expressed as
\begin{equation}
\label{eq:optAlternative2}
\begin{aligned}	
 \max & \quad \delta^2 \mathbf{x}^T A \mathbf{x}+ \delta \left(  \hat{\mathbf{a}}^T A^T -\hat{c} {\mathbf{1}}^T A \right) \mathbf{x } + \hat{c} {\mathbf{1}}^T \hat{\mathbf{a}} \\
s.t. & \quad x_i \in \{-1,1\} \quad \mbox{for all $i\in \cal{I}$}.
\end{aligned}
\end{equation}
Using Lemma \ref{alg1}, it can be seen that Assumption \ref{ass:concavityCond} and  nonnegativity of entries of $\Lambda$ and $G$ imply that $A=(\Lambda-G)^{-1}= \Lambda^{-1} \sum_{l=0}^\infty \left( G \Lambda^{-1} \right)^l$ is a matrix with nonnegative entries.
Therefore,
 Theorem \ref{theo:boundW_pricing} follows by using the formulation \eqref{eq:optAlternative2}, rewriting the pricing problem of the monopolist in the form of \eqref{eq:QBI1},
  and applying Corollary \ref{cor:boundOnQuadGen}.
  
\subsection*{Proof of Lemma \ref{lem:profits}}
Note that ignoring the network effects is equivalent to assuming that $G=0$. Thus,
the optimal prices for the setting described above, denoted by $\mathbf{p}_0$ and $\mathbf{p}_N$ respectively, are given by
\begin{equation}
\mathbf{p}_0= \frac{{\mathbf a}+c{\mathbf 1}
}{2}
\end{equation}
and 
\begin{equation} \label{eq:pNexp}
{\mathbf p}_N= {\mathbf a} -(\Lambda - G) \left(\Lambda - \frac{G+G^T}{2}\right)^{-1} \frac{{\mathbf a}-c{\mathbf 1}
}{2}.
\end{equation}
By Lemma \ref{lem:postiveConsumption}, under the price vector $\mathbf{p}_N$ all agents purchase a positive amount of the good.
Assumption \ref{ass:all_buy} implies that $a_i>c$ for all  $i\in {\cal I}$.
Thus, under the price vector  $\mathbf{p}_0$,  $a_i$ is greater than the price offered to agent $i$, and   agents still  purchase a positive amount of the good.
The corresponding consumption vectors (denoted by $\mathbf{x}_0$ and $\mathbf{x}_N$) are given by (cf. equation \eqref{eq:eqS_unique})
\begin{equation}
\mathbf{x}_0=(\Lambda - G)^{-1}(\mathbf{a}-\mathbf{p}_0)=(\Lambda - G)^{-1}\frac{{\mathbf a}-c{\mathbf 1}
}{2}
\end{equation}
and
\begin{equation} \label{eq:xNexp}
\mathbf{x}_N=(\Lambda - G)^{-1}(\mathbf{a}-\mathbf{p}_N)=
 \left(\Lambda - \frac{G+G^T}{2}\right)^{-1} \frac{{\mathbf a}-c{\mathbf 1}
}{2}.
\end{equation}
\noindent It follows that
\begin{equation}
\Pi_0 = (\mathbf{p}_0-c\mathbf{1})^T\mathbf{x}_0=\frac{{\mathbf a}-c{\mathbf 1}
}{2}(\Lambda - G)^{-1}\frac{{\mathbf a}-c{\mathbf 1}}{2},
\end{equation}
and if we let $M=\Lambda - G$, equations \eqref{eq:pNexp} and \eqref{eq:xNexp} imply that
\begin{equation} \label{eq:PiNExp}
\begin{aligned}
\Pi_N &= (\mathbf{p}_N -c \mathbf{1})^T \mathbf{x}_N \\
&= 
\left( {\mathbf a} -c \mathbf{1} -M \left(\frac{M+M^T}{2}\right)^{-1} \frac{{\mathbf a}-c{\mathbf 1}
}{2} \right)^T
 \left( \left(\frac{M+M^T}{2}\right)^{-1} \frac{{\mathbf a}-c{\mathbf 1} }{2}
\right) \\
&=
\left(\frac{{\mathbf a}-c{\mathbf 1} }{2}\right)^T
\left( 2 I -M \left(\frac{M+M^T}{2}\right)^{-1}  \right)^T
\left(\frac{M+M^T}{2}\right)^{-1}
 \left(  \frac{{\mathbf a}-c{\mathbf 1} }{2} \right)
 \\
 &=
2 \left(\frac{{\mathbf a}-c{\mathbf 1} }{2}\right)^T
 \left(\frac{M+M^T}{2}\right)^{-1}
  \left(  \frac{{\mathbf a}-c{\mathbf 1} }{2} \right) \\
  &\qquad
  -
  \left(\frac{{\mathbf a}-c{\mathbf 1} }{2}\right)^T
\left(\frac{M+M^T}{2} \right)^{-T}
M^T
  \left(\frac{M+M^T}{2}\right)^{-1}
   \left(  \frac{{\mathbf a}-c{\mathbf 1} }{2} \right)
  \\
\end{aligned}
\end{equation}
Note that for a   matrix $A$ and vector $\mathbf{x}$, $\mathbf{x}^T A \mathbf{x}= \mathbf{x}^T \frac{A+A^T}{2} \mathbf{x}$, thus it follows that
\begin{equation*}
\left(\frac{{\mathbf a}-c{\mathbf 1} }{2}\right)^T
\left(\frac{M+M^T}{2} \right)^{-T}
M^T
  \left(\frac{M+M^T}{2}\right)^{-1}
   \left(  \frac{{\mathbf a}-c{\mathbf 1} }{2} \right)
   =
   \left(\frac{{\mathbf a}-c{\mathbf 1} }{2}\right)^T
   \left(\frac{M+M^T}{2} \right)^{-1}
      \left(  \frac{{\mathbf a}-c{\mathbf 1} }{2} \right).
\end{equation*}
Thus, from \eqref{eq:PiNExp} we can rewrite $\Pi_N$ as
\begin{equation}
\Pi_N
=\left(\frac{{\mathbf a}-c{\mathbf 1} }{2}\right)^T
 \left(\frac{M+M^T}{2}\right)^{-1}
  \left(  \frac{{\mathbf a}-c{\mathbf 1} }{2} \right).
\end{equation}
The claim follows noting that $M= \Lambda-G$.

\subsection*{Proof of Theorem \ref{theo:pricingGain}}
To simplify the notation we denote $\Lambda - G$ by $M$.
Note that by the assumption of the theorem, $M$ is positive definite.
 We  state
  some useful properties of this matrix
  in Lemmas \ref{lem:PDM} and \ref{lem:eigM}, and then prove the claim using these properties. 
  The proofs of these Lemmas can be found at the end of this proof.
\begin{lemma} \label{lem:PDM}
If $M$ is positive definite, then
the following matrices are also positive definite: $M^{-1}$, $ \frac{M+M^T}{2}$, $ \left(\frac{M+M^T}{2}\right)^{-1}$, $\left(\frac{M^{-1}+M^{-T}}{2}\right)$ and $\left(\frac{M^{-1}+M^{-T}}{2}\right)^{-1}$.
\end{lemma}
\begin{lemma} \label{lem:eigM}
Let $M$ be positive definite and 
$\lambda$ be an eigenvalue of $MM^{- T}=(\Lambda-G)(\Lambda-G)^{-T}$,
 then:
(i)  $|\lambda|=1$.
(ii) The spectral radius of  $MM^{- T}+M^T M^{-1}$ is smaller than $2$.
\end{lemma}

Let $\mathbf{v}= \left( \frac{\mathbf{a}-c\mathbf{1}}{2} \right)$.
Lemma \ref{lem:profits} implies that
\begin{equation} \label{eq:bound1}
\frac{\Pi_N}{\Pi_0}  = \frac{\mathbf{v}^T  \left(\Lambda- \frac{G+G^T}{2} \right)^{-1} 
\mathbf{v}
}{\mathbf{v}^T (\Lambda - G)^{-1}  \mathbf{v} } \leq \max_{||\mathbf{x}||=1}
\frac{\mathbf{x}^T  \left(\frac{M+M^T}{2} \right)^{-1} 
\mathbf{x}
}{\mathbf{x}^T M^{-1} \mathbf{x}}
=\max_{||\mathbf{x}||=1}
\frac{\mathbf{x}^T  \left(\frac{M+M^T}{2} \right)^{-1} 
\mathbf{x}
}{\mathbf{x}^T \frac{M^{-1}+M^{-T}}{2}  \mathbf{x}}
,
\end{equation}
and similarly
\begin{equation}\label{eq:bound2}
\frac{\Pi_0}{\Pi_N}  \leq 
\max_{||\mathbf{x}||=1}
\frac{\mathbf{x}^T M^{-1}  \mathbf{x}}{\mathbf{x}^T  \left(\frac{M+M^T}{2} \right)^{-1} 
\mathbf{x}
}=
\max_{||\mathbf{x}||=1}
\frac{\mathbf{x}^T \frac{M^{-1}+M^{-T}}{2}  \mathbf{x}}{\mathbf{x}^T  \left(\frac{M+M^T}{2} \right)^{-1} 
\mathbf{x}
}.
\end{equation}

Since $\frac{M^{-T} + M^{-1}}{2}$ and $\frac{M^{T} + M^{1}}{2}$ 
are symmetric positive definite matrices, the matrices 
$\left(\frac{M^{-T} + M^{-1}}{2}\right)^{1/2}$  and $\left(\frac{M^{T} + M^{1}}{2} \right) ^{1/2}$ are well defined.
Consequently, we obtain
\begin{equation}
\begin{aligned}
\max_{||\mathbf{x}||=1}
\frac{\mathbf{x}^T  \left(\frac{M+M^T}{2} \right)^{-1} 
\mathbf{x}
}{\mathbf{x}^T \frac{M^{-1}+M^{-T}}{2}  \mathbf{x}}
&=
\max_{||\mathbf{x}||=1}
\frac{\mathbf{x}^T  \left(\frac{M+M^T}{2} \right)^{-1} 
\mathbf{x}
}{\mathbf{x}^T \left( \frac{M^{-1}+M^{-T}}{2} \right)^{1/2} \left( \frac{M^{-1}+M^{-T}}{2} \right)^{1/2}  \mathbf{x}}\\
&= \max_{||\mathbf{y}||=1}
{\mathbf{y}^T  \left( \frac{M^{-1}+M^{-T}}{2} \right)^{-1/2} \left(\frac{M+M^T}{2} \right)^{-1} 
\left( \frac{M^{-1}+M^{-T}}{2} \right)^{-1/2} \mathbf{y}
}\\
&= \lambda_{max}\left( 
\left( \frac{M^{-1}+M^{-T}}{2} \right)^{-1/2} \left(\frac{M+M^T}{2} \right)^{-1} 
\left( \frac{M^{-1}+M^{-T}}{2} \right)^{-1/2} 
\right),
\end{aligned}
\end{equation}
where the second  line follows by defining 
$\mathbf{z} \triangleq \left( \frac{M^{-1}+M^{-T}}{2} \right)^{1/2}  \mathbf{x}$, rewriting the first line in terms of $\mathbf{z}$ and setting $\mathbf{y}=\frac{\mathbf{z}}{||\mathbf{z}||}$, and the third line  follows from the  Rayleigh-Ritz Theorem (\cite{hornmatrix}).
Similarly we have,
\begin{equation}
\begin{aligned}
\max_{||\mathbf{x}||=1}
\frac{\mathbf{x}^T \frac{M^{-1}+M^{-T}}{2}  \mathbf{x}}{\mathbf{x}^T  \left(\frac{M+M^T}{2} \right)^{-1} 
\mathbf{x}
}
&=
\lambda_{max}\left( 
\left( \frac{M+M^{T}}{2} \right)^{1/2} \left(\frac{M^{-1}+M^{-T}}{2} \right) 
\left( \frac{M+M^{T}}{2} \right)^{1/2} 
\right).
\end{aligned}
\end{equation}

Note that for a real matrix $A$ and invertible real matrix $B$ the eigenvalues of $A$ and $B^{-1 }A B$ are identical (similarity transformation). Therefore, it follows from the above equations that
\begin{equation}
\begin{aligned}
\max_{||\mathbf{x}||=1}
\frac{\mathbf{x}^T  \left(\frac{M+M^T}{2} \right)^{-1} 
\mathbf{x}
}{\mathbf{x}^T \frac{M^{-1}+M^{-T}}{2}  \mathbf{x}}
&= 
\lambda_{max}\left( 
\left( \frac{M^{-1}+M^{-T}}{2} \right)^{-1} \left(\frac{M+M^T}{2} \right)^{-1} 
\right) \\
&= 
\lambda_{max} 
\left(\left(\frac{2I+ MM^{-T}+M^TM^{-1}}{4} \right)^{-1} \right). \\
\end{aligned}
\end{equation}
Lemma \ref{lem:eigM} implies that the eigenvalues of $MM^{-T}+M^{T}M^{-1}$ are real and belong to $[-2,2]$. Thus, it follows that the eigenvalues of $\left(\frac{2I+ MM^{-T}+M^TM^{-1}}{4} \right)$ are positive and 
\begin{equation}
\lambda_{max} 
\left(\left(\frac{2I+ MM^{-T}+M^TM^{-1}}{4} \right)^{-1} \right) \\
= 
1 / \lambda_{min} 
\left(\left(\frac{2I+ MM^{-T}+M^TM^{-1}}{4} \right) \right).
\end{equation}


Similarly we obtain,
\begin{equation}
\begin{aligned}
\max_{||\mathbf{x}||=1}
\frac{\mathbf{x}^T \frac{M^{-1}+M^{-T}}{2}  \mathbf{x}}{\mathbf{x}^T  \left(\frac{M+M^T}{2} \right)^{-1} 
\mathbf{x}
}
&=
 \lambda_{max}\left( 
\left( \frac{M+M^{T}}{2} \right) \left(\frac{M^{-1}+M^{-T}}{2} \right) 
\right)\\
&=
 \lambda_{max}
\left( \frac{2I+ MM^{-T}+M^{T}M^{-1}}{4} \right)  .
\end{aligned}
\end{equation}
 Thus, it follows from \eqref{eq:bound1} and \eqref{eq:bound2} that
\begin{equation}
 \lambda_{min} 
\left(\frac{2I+ MM^{-T}+M^TM^{-1}}{4} \right)
 \leq 
 \frac{\Pi_0}{\Pi_N}  
 \leq 
  \lambda_{max}
  \left( \frac{2I+ MM^{-T}+M^{T}M^{-1}}{4} \right)  ,
\end{equation}
  or equivalently
\begin{equation}
\frac{1}{2}+  \lambda_{min}\left(\frac{ MM^{-T}+M^TM^{-1}}{4} \right)
 \leq 
 \frac{\Pi_0}{\Pi_N}  
 \leq 
\frac{1}{2}+
  \lambda_{max}
  \left( \frac{ MM^{-T}+M^{T}M^{-1}}{4} \right)  .
\end{equation}
The claim follows
since the eigenvalues of $MM^{-T}+M^{T}M^{-1}$   belong to $[-2,2]$.

\subsection*{Proof of Lemma \ref{lem:PDM}}
Note that since $\mathbf{x}^TM\mathbf{x} = \mathbf{x}^T \frac{M+M^T}{2} \mathbf{x}$, it immediately follows that $ \frac{M+M^T}{2}$ is positive definite. 
 For $\mathbf{y}=M\mathbf{x}$, $\mathbf{x}^T M \mathbf{x}= \mathbf{x}^T M^T \mathbf{x}=\mathbf{y}^T M^{-1}\mathbf{y}$. Thus, it follows that  $\mathbf{y}^T M^{-1} \mathbf{y}= \mathbf{y}^T \frac{M^{-T} + M^{-1}}{2} \mathbf{y}>0$ for all real vectors $\mathbf{y}\neq 0$. 
 Hence, $M^{-1}$ and $\frac{M^{-T} + M^{-1}}{2}$ are also positive definite.

Finally, note that if $A$ is a symmetric positive definite matrix, then so is $A^{-1}$. Therefore, positive definiteness of $ \left(\frac{M+M^T}{2}\right)^{-1}$ and $\left(\frac{M^{-1}+M^{-T}}{2}\right)^{-1}$  follows directly from  the fact that $\frac{M+M^T}{2}$ and $\frac{M^{-1}+M^{-T}}{2}$ are positive definite.

\subsection*{Proof of Lemma \ref{lem:eigM}}
Assume that $\mathbf{x}$ is a left eigenvector of  $MM^{-T}$ corresponding to the eigenvalue $\lambda$, i.e. $\mathbf{x}^T MM^{-T}= \lambda \mathbf{x}^T$. Then, $(\lambda, \mathbf{x})$ satisfies 
$\mathbf{x}^T M = \lambda \mathbf{x}^T M^T$
 or equivalently 
 \begin{equation} \label{eq:reordEig}
  M^T\mathbf{x}= \lambda  M\mathbf{x}
 \end{equation}
Since  $MM^{-T}$ need not be a symmetric matrix,  $\lambda$ and $\mathbf{x}$ are not necessarily real. Let $x= x_1 + i x_2$, and $\mathbf{x}^*$ denote the conjugate transpose of $\mathbf{x}$, i.e., $\mathbf{x}^* = x_1^T - i x_2^T$. Note that
\begin{equation}
\mathbf{x}^* M^T\mathbf{x}= x_1^T M^T x_1 + x_2^T M^T x_2  + i\left( x_1^T M^T x_2 - x_2^T M^T x_1 \right),
\end{equation}
and 
\begin{equation}
\mathbf{x}^* M\mathbf{x}= x_1^T M x_1 + x_2^T M x_2  + i\left( x_1^T M x_2 - x_2^T M x_1 \right).
\end{equation}
Since $M$ and $M^T$ are real and positive definite $\Re({\mathbf{x}^* M\mathbf{x}})=\Re({\mathbf{x}^* M^T \mathbf{x}})=x_1^T M x_1 + x_2^T M x_2 >0$. Additionally, taking the transpose, it can be seen that $x_1^T M^T x_2 = x_2^T M x_1$ and $x_1^T M x_2 = x_2^T M^T x_1$ and consequently $\Im({\mathbf{x}^* M\mathbf{x}})=-\Im({\mathbf{x}^* M^T \mathbf{x}})$. Thus from \eqref{eq:reordEig} it follows that
\begin{equation}
|\lambda|= \left| \frac{\mathbf{x}^* M^T \mathbf{x}}{\mathbf{x}^* M \mathbf{x}}  \right| =1.
\end{equation}

Let the Jordan normal form of $MM^{-T}$ be $P^{-1} J P$, i.e.,
\begin{equation}
M M^{-T}=P^{-1} J P,
\end{equation}
where J is an upper triangular block diagonal matrix, and  $P$ is an invertible matrix. Eigenvalues of $M M^{-T}$ correspond to the diagonal entries of $J$. Observing that $(M M^{-T})^{-1}= M^T M^{-1}$, it follows that
\begin{equation}
M M^{-T}+ M^T M^{-1} = P^{-1} J P + P^{-1} J^{-1} P= P^{-1} (J+J^{-1}) P.
\end{equation}
The inverse of an upper triangular block diagonal matrix is upper triangular block diagonal. Thus, it follows that $P^{-1} (J+J^{-1}) P$ is a Jordan normal form for $M M^{-T}+ M^T M^{-1}$. Also note that since $J$ and $J^{-1}$ are upper triangular, if $J_{ii}=\lambda$, then $J^{-1}_{ii}= \frac{1}{\lambda}$. Consequently, the diagonal entries of $(J+J^{-1})$ take the form $\lambda+\frac{1}{\lambda}$, where $\lambda$ is a diagonal entry of $J$. Since $P^{-1} J P$  and $ P^{-1} J P + P^{-1} J^{-1} P= P^{-1} (J+J^{-1}) P$ are the Jordan normal forms of $M M^{-T}$ and $M M^{-T}+ M^T M^{-1}$, we conclude that $\lambda$ is an eigenvalue of $M M^{-T}$ if and only if $\lambda+\frac{1}{\lambda}$ is an eigenvalue of  $M M^{-T}+ M^T M^{-1}$.

From part (i), it follows that the eigenvalues of $M M^{-T}$ take the form  $e^{i\omega}$, for some $\omega \in [0,2\pi)$. Thus, the eigenvalues of $M M^{-T}+ M^T M^{-1}$ are given by $e^{i\omega}+e^{-i\omega}=2 \cos (\omega)$ for some $\omega \in [0,2\pi)$. 
Since $|\cos(\omega)| \leq 1$,
we conclude that the spectral radius (largest eigenvalue in absolute value) of  $M M^{-T}+ M^T M^{-1}$ is bounded by $2$, and the claim follows.

\end{document}